\documentclass[reprint,aps,prl,superscriptaddress,
amsmath,amssymb]{revtex4-1}
\usepackage{mathtools}
\usepackage{graphicx}
\usepackage{booktabs,multirow}
\usepackage{braket}
\usepackage{textcomp}
\usepackage{yfonts}
\usepackage{comment}
\usepackage[
  bookmarks=false,
  colorlinks,
  linkcolor=blue,
  urlcolor=blue,
  citecolor=blue,
  plainpages=false,
  pdfpagelabels,
  final,
  breaklinks=true
]{hyperref}

\begin{document}

\newcommand{\pd}[2]{\frac{\partial #1}{\partial #2}} 
\newcommand{\td}[2]{\frac{d #1}{d #2}} 

\newcommand{\bs}{\boldsymbol}
\newcommand{\bt}{\textbf}
\newcommand{\sech}{\text{sech}}
\newcommand{\erfc}{\text{erfc}}
\newcommand{\bse}{\begin{subequations}}
\newcommand{\ese}{\end{subequations}}
\newcommand{\im}{\text{i}}
\newcommand{\ud}[0]{\mathrm{d}}
\newcommand{\norm}[1]{\left\lVert#1\right\rVert}
\newcommand{\ve}{\alpha}
\newcommand{\op}{\widehat}

\graphicspath{{figures-IG-BH/},{../figures-IG-BH/},{../figures/}} 
\allowdisplaybreaks

\title{Exactly solvable model behind Bose-Hubbard dimers, Ince-Gauss beams, and aberrated optical cavities
}

\author{R. Guti\'errez-Cuevas}
\email{rodrigo.gutierrez-cuevas@espci.fr}
\affiliation{Institut Langevin, ESPCI Paris, Université PSL, CNRS, 
75005 Paris, France}
\affiliation{Aix Marseille Univ, CNRS, Centrale Marseille, Institut Fresnel,
 UMR 7249, 13397 Marseille Cedex 20, France}
\author{D.~H.~J.~O'Dell}
\email{dodell@mcmaster.ca}
\affiliation{Department of Physics and Astronomy, McMaster University, 
1280 Main St. W., Hamilton, Ontario, L8S 4M1, Canada}
\author{M.~R.~Dennis}
\email{m.r.dennis@bham.ac.uk}
\affiliation{School of Physics and Astronomy, University of Birmingham, 
Birmingham B15 2TT, UK}
\author{M.~A.~Alonso}
\email{miguel.alonso@fresnel.fr}
\affiliation{Aix Marseille Univ, CNRS, Centrale Marseille, Institut Fresnel, 
UMR 7249, 13397 Marseille Cedex 20, France}
\affiliation{The Institute of Optics, University of Rochester, Rochester, NY 14627, USA}
\affiliation{Laboratory for Laser Energetics, University of 
Rochester, Rochester, NY 14627, USA}

\date{\today}

\begin{abstract}
By studying the effects of quadratic anisotropy and quartic perturbations 
on the two-dimensional harmonic oscillator, one arrives at 
a simple model termed here the Ince oscillator, 
whose analytic solutions are given in terms of Ince polynomials. 
This one model unifies diverse physical systems, 
including aberrated optical cavities that are shown to support Ince-Gauss 
beams as their modes, and the two-mode Bose-Hubbard dimer describing two 
coupled superfluids. The Ince oscillator model describes a topological 
transition which can have very different origins: in the optical case, 
which is fundamentally linear, it is driven by the ratio of astigmatic 
to spherical mirror aberrations, whereas in the superfluid case it is 
driven by the ratio of particle tunneling to interparticle interactions 
and corresponds to macroscopic quantum self trapping.
\end{abstract}

\maketitle

{\it Introduction}. 
Analogies between physical phenomena arise when, in certain limits, their different fundamental equations reduce to similar models. 
A very basic model is the harmonic oscillator (HO), which in both path/classical or wave/quantum forms describes a variety of phenomena in mechanics and optics
\cite{sakurai2010modern,siegman1986lasers}. 
The isotropic 2-dimensional HO (2DHO) is more striking, since it possesses a hidden 
SU(2) symmetry that endows it with three constants of the motion (CoMs) 
involving its position  and momentum. 
For the classical case these constants take the form
\begin{align}
   L_j \equiv \frac{1}{4}\left(\frac{\mathbf{q}}{\sqrt{\gamma}}-\im\sqrt{\gamma}
   \mathbf{p}\right)\sigma_j \left(\frac{\mathbf{q}}{\sqrt{\gamma}}
   +\im\sqrt{\gamma}\mathbf{p}\right),
\end{align}
for $j = 1,2,3$, where ${\bf q}=({q}_x,{q}_y)$ is the position, 
${\bf p}=({p}_x,{p}_y)$ is the momentum, $\gamma$ is a positive constant 
with the appropriate 
units for the problem in question, and $\sigma_j$ are the (permuted) Pauli matrices  
\begin{align}
\sigma_1&=\left(\begin{array}{cc}1&0\\0&-1\end{array}\right),\,\,
\sigma_2=\left(\begin{array}{cc}0&1\\1&0\end{array}\right),\,\,
\sigma_3=\left(\begin{array}{cc}0&-\im\\\im&0\end{array}\right).
\end{align}
Note that $L_1^2+L_2^2+L_3^2 = (\tfrac{1}{4\gamma}q^2+\tfrac{\gamma}{4}p^2)^2$ 
is proportional to the Hamiltonian squared, where 
$q^2=\mathbf{q}\cdot\mathbf{q}$ and $p^2=\mathbf{p}\cdot\mathbf{p}$. 
The Poisson brackets (or the commutators in the quantum case) of the CoMs 
resemble angular momentum algebra. 
Of these CoMs, $L_1$ and $L_2$ are quadratic in both ${\bf p}$ and ${\bf q}$, 
but $L_3 =( q_x p_y-q_y p_x)/2$ [half the orbital angular momentum (OAM)] is linear in both 
${\bf p}$ and ${\bf q}$, the linearity in ${\bf p}$ playing a central role 
in what follows. 

Here, we present an analytically solvable, nonlinear model we call the 
\emph{Ince oscillator} based on the quantum 2DHO experiencing 
a perturbation whose effect is described by the Ince operator $\widehat{\mathcal{I}}$, 
defined as
\begin{equation}
   \widehat{\mathcal{I}} =  \frac{\alpha}2 \widehat{L}_1+\widehat{L}_3^2 ,
   \label{eq:inceop}
\end{equation}
where $\widehat{L}_j$ are the Fradkin-Stokes operators corresponding to the SU(2) 
CoMs, and $\alpha$ is a positive parameter.
This system can be visualized physically as a 2DHO subject to the two simplest 
meaningful perturbations: 
quadratic asymmetry and a quartic correction to the potential (determined by 
$\widehat{L}_1$ and $\widehat{L}_3^2$ respectively). 
It can then be considered as the simplest generalization of a 2DHO that not 
only presents topological transitions, but can also be solved analytically, 
in terms of Ince polynomials 
\cite{ince1925linear,arscott1964periodic}.
This simplicity endows it with an important level of universality, 
so that it describes a range of different physical systems. 
We emphasize two such systems for which the connection is 
particularly surprising. 
The first is a linear, optical resonant cavity like those used in lasers and 
high-precision interferometers \cite{abbot2016observation,tao2020higher}. In 
the presence of small amounts of simple aberrations, the cavity modes are shown to correspond to Ince-Gauss (IG) beams, 
which have received significant attention recently  
\cite{ince1925linear,arscott1964periodic,boyer1975liea,boyer1975lie,
bandres2004incegaussiana,bandres2004incegaussian, schwarz2004observation,
krenn2013entangled,plick2013quantum,woerdemann2011optical}. 
The second system is the Bose-Hubbard (BH) dimer model 
\cite{milburn1997quantum,smerzi1997quantum,raghavan1999coherent,anglin2001exact,leggett2001bose,gati2007bosonic,graefe2007semiclassical,sakmann2009exact,nissen2010wentzel,chuchem2010quantum,odell2012quantum,graefe2014bosehubbard}, 
a workhorse in condensed matter physics providing
a minimal model for coupled 
reservoirs of superfluid helium 
\cite{sukhatme2001observation,backhaus1998discovery}, coupled atomic 
Bose-Einstein condensates (BECs) 
\cite{cataliotti2001josephson,albiez2005direct,schumm2005matter,
levy2007a.c.,
leblanc2011dynamics,gerving2012non,trenkwalder2016quantum}, and coupled 
polariton BECs in semiconductor microcavities 
\cite{lagoudakis2010coherent,abbarchi2013macroscopic}, among others.

{\it 2D harmonic oscillator}. 
The 2DHO obeys the
Schr\"odinger equation 
$\im\eta\,\partial_\tau\ket{\psi}=\op{H}_0\ket{\psi}$, 
where $\eta$ is a constant (e.g. the reduced Planck constant for 
mechanics or the reduced wavelength for optics), $\tau$ is the 
propagation/evolution parameter,  
and $\op{H}_0$ is the Hamiltonian 
\begin{align}
\label{eq:HOT}
    \op{H}_0 = \frac{\kappa}{2} \left( \frac{1}{\gamma}\op{q}^2
    +\gamma \op{p}^2 \right),
\end{align}
where $\kappa$ is also a constant with appropriate 
units for the problem in question.
In the position representation, $\op{\bf q}\rightarrow (q_x,q_y)$ and $ \op{\bf p} \rightarrow 
-\im \eta (\partial_{q_x},\partial_{q_y})$. The eigenvalues of the 
Hamiltonian in Eq.~(\ref{eq:HOT}) 
are $\kappa\eta (N+1)$, 
where $N$ is a non-negative integer referred to as the total order. There are $N+1$ degenerate 
eigenstates of $\op{H}_0$ for any $N$, so the choice of a set of 
eigenfunctions is not unique 
\cite{sakurai2010modern,gutierrez-cuevas2020modal}, and different eigenstates are obtained through separation of variables in several ways \cite{boyer1975liea,boyer1975lie}. 
For example, separation in Cartesian coordinates
leads to Hermite-Gauss 
(HG) modes, while separation in polar coordinates gives 
Laguerre-Gauss (LG) modes \cite{siegman1986lasers}.

Schwinger's  coupled oscillator model \cite{sakurai2010modern} provides an elegant description of the 2DHO, based on the Fradkin-Stokes operators 
$\op{\bt L}\equiv(\op L_1, \op L_2, \op L_3)$,
which 
satisfy the $\mathfrak{su}(2)$ commutation relations 
 $ [\op L_i,\op L_j]= \im \eta\sum_k\epsilon_{ijk} 
\op L_k$,
with   $ \epsilon_{ijk}$ being the Levi-Civita tensor.
These operators commute 
with the unperturbed Hamiltonian $[\op{H}_0,\op L_j]=0$ since
\begin{align}
    \op{\bf L}\cdot\op{\bf L}=\op{L}_1^2+\op{L}_2^2+\op{L}_3^2
    =\frac{1}{4\kappa^2}\op{H}_0^2-\frac{\eta^2}4.
\label{eq:Tsquared}
\end{align}
Thus, the degenerate set 
of modes of $\op{H}_0$ with equal $N$ can be mapped onto a 
collective spin with total angular momentum 
$N/2$ 
\cite{dennis2017swings,gutierrez-cuevas2019generalized,gutierrez-cuevas2020modal,
dennis2019gaussian,
gutierrez-cuevas2020platonic}.
Different spin axes correspond to different modes 
\cite{dennis2017swings,gutierrez-cuevas2019generalized,
gutierrez-cuevas2020modal}: 
the HG and LG modes are eigenstates 
of $\op L_1$ and $\op L_3$, respectively. 

{\it Perturbed 2DHO}. 
The degeneracy of the 2DHO can be removed by adding a small 
perturbation $\op{W}$ to the Hamiltonian:
\begin{align}
    \op{H}= \op{H}_0 +
    \op{W}(\op{\bf q},\op{\bf p}),
\end{align}
Propagation/evolution over an interval $\tau$ is described by the operator
$\exp(-\im\tau\op{H}/\eta) = 
  \exp[-\im\tau(\op{H}_0+ \op{W})/\eta]$.
Since 
$\op{W}$ is small, we can use the first Born 
approximation to arrive at
\begin{align}
    \exp(-\im\tau\op{H}/\eta)
    &=\exp(-\im\tau\op{H}_0/\eta)\left(1-\frac{\im \tau}{\eta}\op{\cal P}\right),
\end{align}
where 
\begin{align}
    \op{\cal P}&=\frac{1}{\tau}\int_0^\tau\ud\tau'\exp\left(\frac{\im\tau'}{\eta}\op{H}_0\right)\,
    \op{W}(\op{\bt q}, \op{\bt p})\exp\left(-\frac{\im\tau'}{\eta}\op{H}_0\right).
\end{align}
By assuming that $\tau$ is much larger than the oscillation period, 
changing variables to $\kappa\tau'=\theta$, and using the canonical
commutation relation,
we can write this operator as
\begin{align}
\label{eq:Pav}
    \op{\cal P}
    &\approx\frac1{2\pi}\int_0^{2\pi}\ud\theta
    \op{W}\left(\op{\bf q}\cos\theta+\gamma\op{\bf p}\sin\theta,
    \op{\bf p}\cos\theta-\op{\bf q}\frac{\sin\theta}\gamma\right).
\end{align}
The operator $\op{\cal P}$ is then simply the accumulation of the 
effect of $\op{W}$ over many cycles. This accumulation has the 
effect of making $\op{\cal P}$ symmetric in 
$\op{\bf q}$ and $\op{\bf p}$, so that it commutes with the unperturbed 
Hamiltonian $\op{H}_0$. Since we are in the perturbative regime, the 
modes must then be eigenstates of both $\op{H}_0$ and $\op{\cal P}$. 

We consider simple forms for $\op{W}$, corresponding to 
low-order polynomials in position and momentum. It is 
easy to show that Eq.~(\ref{eq:Pav}) vanishes for any odd-order monomial, 
so it is 
sufficient to consider even powers. 
Since we consider 
simple physical systems where the kinetic and potential parts are 
separate, we focus on even-order monomials that include either $\op{\bf q}$ or $\op{\bf p}$.   

\emph{Ince oscillator.} We define the Ince oscillator as a 2DHO subject 
to the two simplest perturbations: an anisotropic quadratic 
and a rotationally symmetric quartic terms. Therefore, we set 
\begin{align} \label{eq:Ws}
  \op{W}=\frac{\epsilon_1}{2\gamma}\left(\op{q}_y^2-\op{q}_x^2\right)
  +\frac{\epsilon_2\op{q}^4}{\gamma^2}
\end{align}
where the coefficients $\epsilon_1$ and $\epsilon_2$ are assumed to be small, justifying the perturbative approach. This leads to
\begin{align}
  \op{\cal P}
  &=   2 \epsilon_2\left(\frac{3}{4\kappa^2}\op{H}_0^2
  +\frac{\eta^2}{4}- \op{\mathcal{I}} \right).
  \label{eq:PInce}
\end{align}
where $\op{\mathcal{I}}$ is the Ince operator defined in Eq.~(\ref{eq:inceop}), with 
$\alpha/\eta = \epsilon_1/ \eta \epsilon_2$ being a dimensionless parameter that can take any real value.
Note that if we had replaced the position operators in Eq.~(\ref{eq:Ws}) with the 
corresponding momentum operators (with the appropriate factors of $\gamma$ for dimensional reasons) the result would have been the same.
In particular, a quartic perturbation in momentum would correspond to
a relativistic correction for quantum-mechanical particle evolution or a postparaxial correction for monochromatic optical propagation.
The two terms in the perturbation each break the degeneracy in a different way, hence selecting a specific 
set of modes: HG modes (separable in Cartesian coordinates) are 
eigenstates of $\op L_1$ so they are selected by quadratic asymmetry 
($\alpha/\eta \rightarrow \infty$),
while LG modes (separable in polar coordinates) are eigenstates of 
$\op L_3$ 
and are hence selected by a rotationally-symmetric 
quartic perturbation ($\alpha/\eta=0$). Since $\op{\cal P}$ 
is quadratic in $\op L_3$, some degeneracy remains in this case between LG modes with the same amplitude profile but opposite OAM.

\begin{figure}
  \centering
  \includegraphics[width=.99\linewidth]{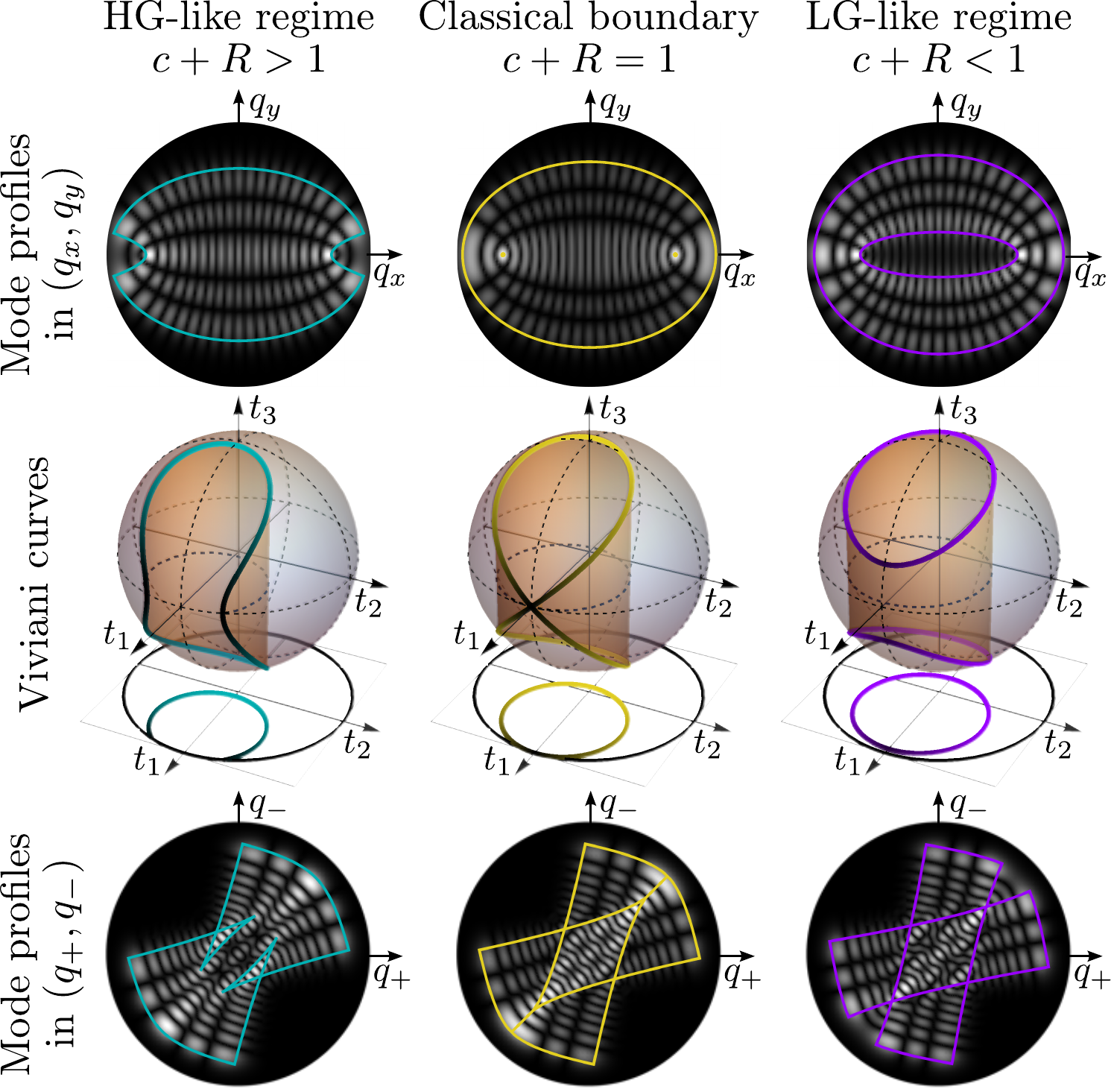}
  \caption{\label{fig:raysIG} Various representations of the 
  topological transition in the classical 
  limit for the same even IG mode with $N=22$ 
  and $\mu=18$. The left and right columns correspond, 
  respectively, to the two topologically distinct 
  regimes: HG-like (turquoise), and LG-like (purple); 
  the middle column shows the
  boundary between them (yellow). 
  The first row depicts 
  $|\braket{{\bf q}|\psi_{N,\mu}^{(p)}(\ve )}|$ (corresponding, e.g., 
  to the transverse profile of an IG beam)
  and whose significant values are contained within the
  caustics, indicated by the overlaid color curves.
  The second row shows the generalized Viviani curves on the 
  Bloch/Poincar\'e sphere.
  The third row shows $|\langle q_+,q_-|\psi_{N,\mu}^{(p)}(\ve )\rangle|$, the amplitude over the real parts of the quadratures of two modes in a BH dimer, as well as the corresponding caustics. Note that $\langle q_+,q_-|\psi_{N,\mu}^{(p)}(\ve )\rangle$ can be calculated from $\braket{{\bf q}|\psi_{N,\mu}^{(p)}(\ve )}$ by Fourier transforming the latter in $q_y$ and rotating the result by $\pi/4$.
  }
\end{figure}

For an eigenstate of $\op{H}_0$ to also be 
an eigenstate of $\op{\cal P}$, it is sufficient that it be 
an eigenstate of $\op{\cal I}$:
\begin{align} 
\label{eq:eigI}
\op{\mathcal{I}} \; \ket{\psi_{N,\mu}^{(p)}(\ve )} 
=& \frac{\eta^2 a_{N,\mu}^{(p)}}{4} \ket{\psi_{N,\mu}^{(p)}(\ve )},
\end{align} 
where 
each mode is identified by its total order $N$, 
parity $p$, and index $\mu$, ordered such that the corresponding 
eigenvalues satisfy $a^{(p)}_{N,\mu } <a^{(p)}_{N,\mu +2 } $ and 
$a^{(o)}_{N,\mu } <a^{(e)}_{N,\mu} $.
As discussed in the Supplemental Material \cite{SM}, 
\nocite{alonso2011wigner,healy2015linear,lipkin1965validity,
das2006infinite,britton2012engineered,richerme2014non,bohnet2016quantum,
zibold2010classical,esteve2008squeezing,gross2010nonlinear,
julia-diaz2012dynamic,haroutyunyan2003analogy,krahn2009classical,
veksler2015semiclassical,mumford2019quantum,chin2010feshbach,
susskind1964quantum,carruthers1968phase,pegg1989phase}
these eigenstates are the IG modes 
\cite{ince1925linear,arscott1964periodic,boyer1975liea,boyer1975lie,
bandres2004incegaussiana,
bandres2004incegaussian}, which in the position 
representation are separable in elliptical coordinates 
with focal separation equal to 
$2 (\alpha \gamma)^{1/2}$. The dependence on each variable is a combination of 
an exponential and an Ince 
polynomial. 
The spatial profile $|\braket{{\bf q}|\psi_{N,\mu}^{(p)}(\ve )}|$ of some 
of these modes is shown in Fig.~\ref{fig:raysIG}. This profile depends 
on $\ve$: for $\ve/\eta \to\infty$ the modes reduce to HG modes, 
while for $\ve/\eta \to0$ they reduce to {\it real} LG modes 
(the superposition 
in equal amounts of two LG modes of equal radial profile and opposite 
vorticity). For intermediate values of $\ve$ the modes resemble 
deformed versions of either HG 
or real LG modes; we refer to these as the HG-like and LG-like regimes, respectively. 
We emphasize that although our derivation of the equations 
defining the Ince oscillator involves perturbation theory, our treatment 
of the Ince operator itself [Eq.~(\ref{eq:eigI})] is exact.


\emph{Classical limit.} In order to better understand the two  
regimes for the IG modes and the transition between them, 
consider the ``classical'' limit (valid for large $N$) obtained by 
replacing the 
operators with c-number quantities. The classical version of 
Eq.~(\ref{eq:Tsquared}) can be written as $t_1^2 + t_2^2 +t_3^2 = 1$. 
where $t_j=2L_j/(\eta(N+1))$ 
This relation defines a unit Poincar\'e/Bloch sphere 
in the space ${\bf t}=(t_1,t_2,t_3)$. 
However, the modes must also be eigenstates of the Ince operator. 
The classical version of 
Eq.~(\ref{eq:eigI}) can be written as 
\begin{align}
  t_3^2+\ve t_1 /\eta (N+1) = a/(N+1)^2,
\end{align}
 where $a$ is the eigenvalue. 
This equation corresponds to a parabolic cylinder in the space 
of ${\bf t}$. The mode can then be represented by the intersection 
of the unit sphere and this parabolic cylinder. 
 This intersection is equivalent \cite{graefe2014bosehubbard} to that of 
the sphere with the vertical cylinder 
$\left(t_1 -c \right)^2 +t_2^2 = R^2$,
aligned along the $t_3$ axis, centered at $c=\ve /2\eta(N+1)$ and of radius 
$R=\{ 1  +[(\ve/2\eta)^2-a]/(N+1)^2 \}^{1/2}$ as shown in 
Fig.~\ref{fig:raysIG}. These 
curves correspond to generalized Viviani curves, also known as Euclid 
spherical ellipses.

The generalized Viviani curves present two topologically distinct 
regimes, each 
linked to a type of IG mode. 
When 
$c+R>1$ (left-hand column of Fig.~\ref{fig:raysIG}), 
the intersection of the sphere and the cylinder is composed 
of a single loop. The projection of this intersection onto the 
$(t_1,t_2)$ plane is an open circular segment that can be used to determine 
the caustics of the mode \cite{alonso2017ray,dennis2017swings}, 
which tend to coincide with the inflection points of the mode's 
amplitude at the edges of the areas occupied by the modes. 
In this case, the caustics take the shape of a curvilinear quadrangle composed of
two sections of an ellipse and a section of each of the two branches of 
a confocal hyperbola. This shape mimics the profile of 
the HG-like mode.  
On the other hand, when $c+R<1$ (right-hand column of Fig.~\ref{fig:raysIG}) 
the intersection between the cylinder 
and the sphere is composed 
of two loops, both of which project onto a closed circle in the 
$(t_1,t_2)$ plane, from which the caustics can be found to be two 
complete confocal ellipses, so that the mode resembles an elongated 
real LG mode.
The boundary case $c+R=1$ (middle column of Fig.~\ref{fig:raysIG}), 
where the topological transition takes place, 
corresponds to a figure of 
eight-shaped curve (the original Viviani curve being a special case). 

\begin{figure}
\centering
\includegraphics[width=.9\linewidth]{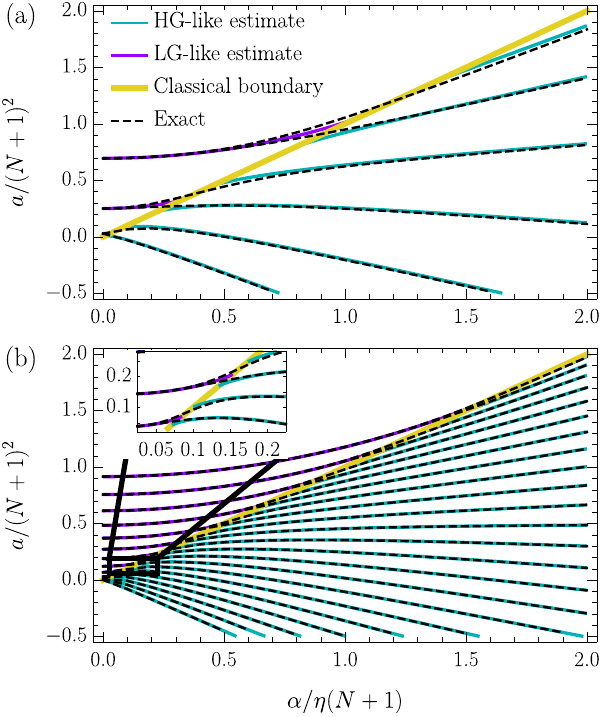}
\caption{\label{fig:eigen} Semiclassical and exact eigenvalues for the 
IG modes with (a) $N=5$ and (b) $N=22$, as functions of $\ve$. }
\end{figure}

The topological transition is not only manifested in 
the shape of the eigenmodes, but also in the near-degeneracy of the 
corresponding eigenvalues as shown in Fig.~\ref{fig:eigen}. In the 
semiclassical limit, the eigenvalues can be determined through
self-consistency conditions in the wave estimates 
\cite{graefe2007semiclassical,shchesnovich2008fock,nissen2010wentzel,
odell2012quantum,alonso2017ray,dennis2019gaussian,
zor1996globally,forbes2001using,alonso2001using,alonso2002stable} in which 
the solid angle enclosed by each loop of the generalized 
Viviani curves must be quantized as an odd multiple of $2\pi/(N+1)$ \cite{alonso2017ray}. 
This quantization is related to a geometric phase through 
the Pancharatnam-Berry connection 
\cite{pancharatnam1956generalized,berry1987adiabatic,
malhotra2018measuring,dennis2019gaussian}. In the  HG-like regime, where 
the curve consists of only one loop, the total subtended solid angle must 
be an odd multiple of $2\pi/(N+1)$, but in the
LG-like regime, where the curve is  composed of two loops, each loop must 
satisfy this condition so the total solid angle must be an even multiple 
of $2\pi/(N+1)$. This discrepancy creates a discontinuity at the topological transition at which the semiclassical estimate fails, as shown in 
Fig.~\ref{fig:eigen}. Away from the transition, the 
semiclassical estimate for the eigenvalue in Eq.~(\ref{eq:eigI}) is 
nearly indistinguishable from the exact eigenvalue, even for small $N$.

{\it Optical cavities}. 
The first physical realization of the Ince oscillator discussed here
is a linear (non-interacting) system corresponding to an aberrated optical cavity. 
Optical cavities    
are a central component of laser systems,
and their shape determines the transverse profile of the
generated beam 
\cite{siegman1986lasers,shen2020structured}.
Cavities are
also used
in high-precision interferometric measurements
\cite{abbot2016observation}. 
These 
applications have motivated studies of
the effect of optical aberrations 
\cite{jaffe2021aberrated} on the cavity modes. For instance, 
particular structured modes have been shown to be resilient to 
small amounts of aberrations in the cavities used
for interferometric gravitational wave detection
\cite{tao2020higher}, and nonplanar cavities have been used to produce 
Laughlin light states 
\cite{schine2016synthetic,clark2020observation}. However, very few 
cases lead to solutions in terms of simple 
closed-form expressions. 

\begin{figure}
  \centering
  \includegraphics[width=.99\linewidth]{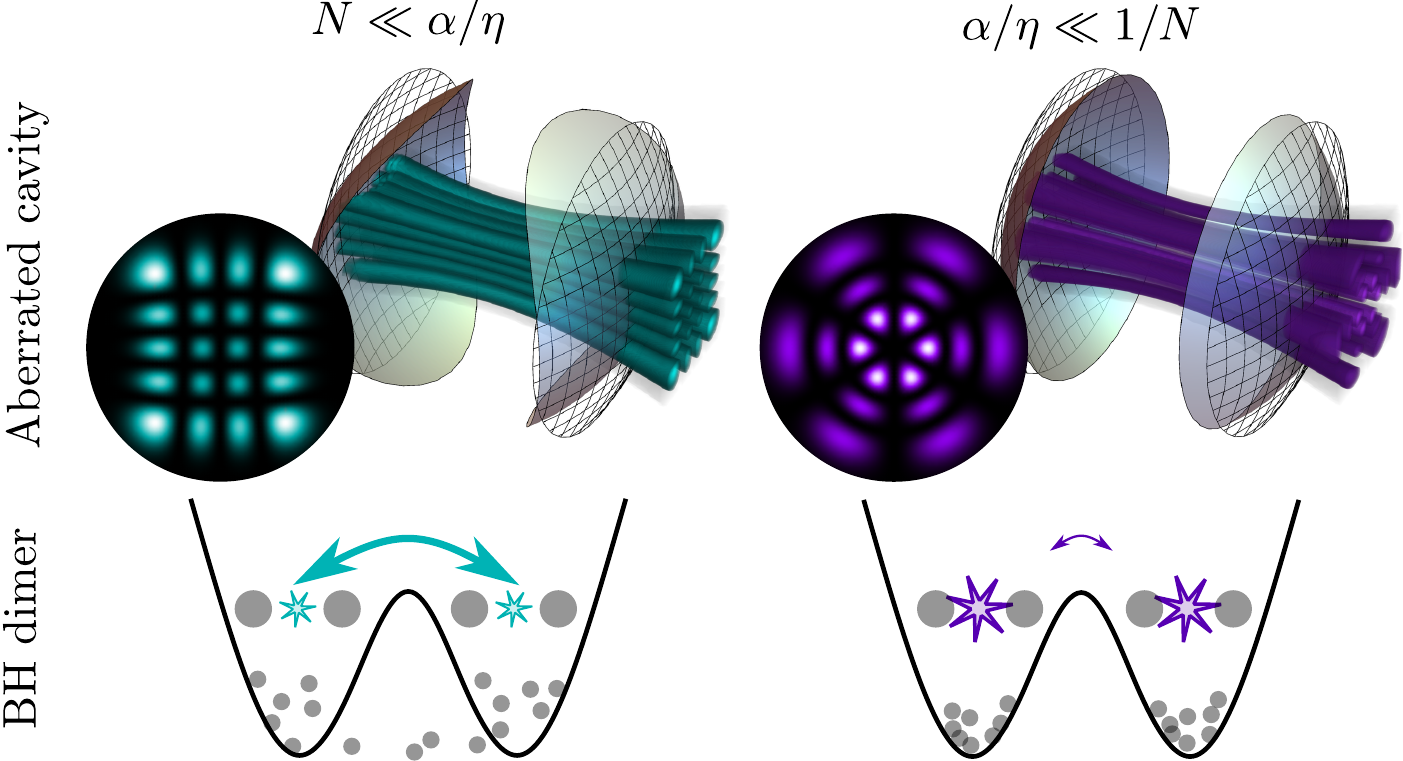}
  \caption{\label{fig:sum} Physical realizations. Top row: 
  IG modes of total order $N$ supported by an optical 
  cavity with small amounts of astigmatism and spherical aberration, 
  where the round insets show the transverse intensity profiles;
  Bottom row: representation of the BH dimer model as a
  BEC with $N$ particles in a double well potential subject to on-site
  interactions (represented by stars) and hopping (represented by double arrows).
  For the BH dimer $\ve$ represents the ratio between the two types of interaction, 
  while for the IG modes it is the ratio
  between the two types of aberration in the cavity.
 Each system is depicted for values of the parameters well into one of the two regimes: (turquoise) Rabi/HG-like regime where 
  hopping/astigmatism dominate,
  and
  (purple) Fock/LG-like regime where on-site interactions/spherical aberration 
  dominate.
  The figure depicts the case
  of $N=7$, even, $\mu=3$, (left) $\alpha/\eta=5N$, and (right) $\alpha/\eta=1/5N$.}
\end{figure}

In the paraxial regime, a resonant cavity composed of two identical 
unaberrated curved mirrors can be mapped onto a 2DHO
\cite{boyer1975liea,boyer1975lie,siegman1986lasers,dennis2017swings,
dennis2019gaussian,gutierrez-cuevas2020modal}. 
The cavity is perturbed if the mirrors present slight aberrations 
that deform the wavefront after each reflection. Two of the simplest 
optical aberrations are (Fig.~\ref{fig:sum}): 
astigmatism, introduced by a 
deviation from rotational symmetry in the shape of the mirrors or by 
a slight misalignment of the system; and spherical aberration, 
which is a quartic deviation of the mirror's ideal radial shape. 
(A small deviation from paraxiality 
has a similar effect as spherical aberration.)
A slight difference
between the 2DHO and the cavity is that for the latter both the main quadratic 
potential and the aberrations
act discretely each time the beam is reflected by the mirrors. However,
it is shown in the Supplemental Material \cite{SM} that for a stable 
non-confocal cavity with a high quality factor, 
an averaging effect like that in Eq.~(\ref{eq:Pav}) takes place
so that the resulting 
modes are eigenstates not only of $\widehat{H}_0$ but also of 
$\op{\cal I}$, where now $\ve$ quantifies the ratio between astigmatism 
and spherical aberration. They 
are therefore IG modes, 
which resemble
HG modes when 
astigmatism dominates ($\ve/\eta \rightarrow \infty$) and LG modes when 
spherical 
aberration dominates ($\ve/\eta \rightarrow 0$). 
When propagating outside the cavity, IG modes are referred to as IG beams, 
whose applications have included micro-manipulation 
\cite{woerdemann2011optical}, encoding of quantum information 
\cite{krenn2013entangled,plick2013quantum}, and studies of vortex 
breakup 
\cite{dennis2006rows}.
It has been observed experimentally that IG beams result from resonators 
with slightly tilted or shifted mirrors
\cite{schwarz2004observation}, and earlier theoretical and experimental 
studies of imperfect optical cavities (without the use of the paraxial 
approximation) \cite{chao1974high} found modes whose transverse profiles 
in retrospect resemble those of IG modes. The analysis presented 
here provides a rigorous foundation for this connection, where a clear 
relation is given between the parameters for the shape of the IG mode 
and those of the cavity.

 {\it BH dimer}. 
Let us now consider the second-quantized form of the Fradkin-Stokes operators 
in terms of the annihilation operators 
$\op a_j =(\op q_j +\im\,\op p_j)/2^{1/2}$ with $j=x,y$ (setting $\eta=\gamma=1$ 
for simplicity), and perform the canonical transformation 
$\op a_{\pm} = (\op a_x \mp \im \op a_y)/2^{1/2}$. The Ince operator can then be 
identified with the two-mode BH Hamiltonian, where 
$\op L_1 = (\op{a}_+^\dag \op{a}_-+ \op{a}_
+ \op{a}_-^{\dag})/2$ describes particle hopping and $\op L_3 = 
(\op{a}_+^{\dag} \op{a}_+- \op{a}_-^{\dag} \op{a}_-)/2$ describes the particle 
number difference between
the two modes.
The term with $\op L_3^2$ accounts for on-site 
particle-particle interactions. Here, $\alpha$ controls the behavior of the system 
\cite{leggett2001bose,gati2007bosonic} by fixing the 
ratio between the hopping and the interactions, and
$N$ corresponds to the total number of particles in the system.
When $N \ll \ve/\eta$  (the Rabi regime) particles can hop freely between 
sites, 
while in the opposite limit $\ve/\eta \ll 1/N$ (the Fock regime) the 
population difference is 
locked by strong interactions, a phenomenon known as macroscopic 
quantum self trapping (MQST) \cite{raghavan1999coherent} 
which is a manifestation of the topological transition on the 
Bloch sphere (see the Supplemental material \cite{SM} for more information
about the BH dimer).
Due to the canonical transformation, the two modes of the BH dimer do not correspond 
to the modes of the coordinates axes of the 2DHO, but rather 
to the two modes defining the sign of the OAM of the 2DHO. 
Additionally, the particle number difference between the two modes of 
the BH dimer corresponds to the net OAM in the Ince oscillator.
It is then possible to write the representation of the wavefunction in terms of 
the real parts of the quadratures for each site from the corresponding IG modes, 
namely
$\langle q_+,q_-|\psi_{N,\mu}^{(p)}(\ve )\rangle$, by applying the operators $ \exp(-\im\pi \op L_3/2\eta)\exp(\im\pi \op L_1/2\eta)$ to the spatial modes
$\langle{\bf q}|\psi_{N,\mu}^{(p)}(\ve )\rangle$. 
These two unitary operations are rotations of the 
Bloch sphere, which in coordinate space correspond to an antisymmetric 
fractional Fourier transform followed by a physical rotation 
\cite{gutierrez-cuevas2020modal}. 
Thus, the eigenstates of the BH dimer can be written as closed-form solutions of the 
Ince oscillator.  An example of one of these wavefunctions and the corresponding caustics, 
calculated from the projection of the curve onto the $(t_1,t_3)$ plane 
\cite{alonso2017ray,dennis2017swings}, are shown  in Fig.~\ref{fig:raysIG}. 
These naturally two-dimensional solutions form a complete orthogonal basis that avoids 
the subtle issues associated with other approaches to the BH dimer such as the 
Bargmann state representation which is over complete 
\cite{anglin2001exact,ulyanov1992new,mumford2017catastrophes,vourdas1990su2,luis1993phase}.

\emph{Conclusions and outlook}. 
We have shown that the Ince oscillator, namely a 2DHO subject to anisotropic 
and quartic perturbations, corresponds to several apparently unrelated physical 
systems. Examples of such systems not treated explicitly here include the modes 
of slightly anisotropic gradient index waveguides, bundles of coupled waveguides 
\cite{longhi2011optical}, and the evolution of polarization in a birefringent nonlinear 
medium \cite{morales2017polarization}. Here we focused on two systems: an aberrated 
optical cavity and the BH dimer. 
We showed that the aberrated optical cavity has eigenmodes that are 
separable in elliptical coordinates, 
which correspond to the IG beams that have been widely used in structured light 
applications \cite{ince1925linear,arscott1964periodic,boyer1975liea,
boyer1975lie,
bandres2004incegaussiana,
bandres2004incegaussian, schwarz2004observation,krenn2013entangled,
plick2013quantum,woerdemann2011optical}. The classical analysis shows why there are 
two regimes for these modes, and explains the geometry of each. 
Furthermore, identifying the BH dimer as an Ince oscillator shows that
the Ince polynomials are
analytic representations of its eigenstates, a fact that has thus far been overlooked.  
It also shows that any system describable by the Ince oscillator shows the same rich 
dynamics as the bosonic Josephson junction, including analogues of plasma oscillations, 
pi-oscillations and macroscopic quantum self trapping.

The Ince oscillator and its modes have connections yet 
to be explored to several other physical systems, including the planar 
quantum pendulum and the Razavy potential 
\cite{becker2017conditional,razavy1980exactly}. 
Moreover, the classical 
limit of this model leads to even more connections not treated here, such as the non-rigid pendulum \cite{smerzi1997quantum}, the simple pendulum \cite{graefe2014bosehubbard}, and the rotational dynamics of celestial bodies \cite{colombo1966cassinis,henrard1987colombos}.

A number of possible extensions can be considered. The effect 
of other perturbations, such as quartic astigmatism, can be 
studied with the expectation that it would lead to other implementations
of the BH dimer and its generalizations. In particular, 
a slow rotation of the 2DHO would induce a perturbative term proportional to $\op L_3$, which would allow modeling nonplanar ring cavities and  
unbalanced (tilted) BH dimers. Another interesting generalization is the effect of 
perturbations in the 3DHO where there are even more separable families of 
solutions, leading to a model for the more complex dynamics of the three-site BH system
\cite{law1998quantum,evrard2021many,kirkby2022caustics}.
These connections 
and extensions will be explored in future work.

The authors acknowledge J.M. Fellows for useful suggestions.
The work by R.G.C. and M.A.A. was supported by the Excellence Initiative of 
Aix-Marseille University - A*MIDEX, a French ``Investissements d'Avenir'' 
programme. D.H.J.O. was supported by NSERC (Canada). MRD acknowledges support 
from the EPSRC Centre for Doctoral Training in Topological Design (EP/S02297X/1).


\begin{thebibliography}{90}%
  \makeatletter
  \providecommand \@ifxundefined [1]{%
   \@ifx{#1\undefined}
  }%
  \providecommand \@ifnum [1]{%
   \ifnum #1\expandafter \@firstoftwo
   \else \expandafter \@secondoftwo
   \fi
  }%
  \providecommand \@ifx [1]{%
   \ifx #1\expandafter \@firstoftwo
   \else \expandafter \@secondoftwo
   \fi
  }%
  \providecommand \natexlab [1]{#1}%
  \providecommand \enquote  [1]{``#1''}%
  \providecommand \bibnamefont  [1]{#1}%
  \providecommand \bibfnamefont [1]{#1}%
  \providecommand \citenamefont [1]{#1}%
  \providecommand \href@noop [0]{\@secondoftwo}%
  \providecommand \href [0]{\begingroup \@sanitize@url \@href}%
  \providecommand \@href[1]{\@@startlink{#1}\@@href}%
  \providecommand \@@href[1]{\endgroup#1\@@endlink}%
  \providecommand \@sanitize@url [0]{\catcode `\\12\catcode `\$12\catcode
    `\&12\catcode `\#12\catcode `\^12\catcode `\_12\catcode `\%12\relax}%
  \providecommand \@@startlink[1]{}%
  \providecommand \@@endlink[0]{}%
  \providecommand \url  [0]{\begingroup\@sanitize@url \@url }%
  \providecommand \@url [1]{\endgroup\@href {#1}{\urlprefix }}%
  \providecommand \urlprefix  [0]{URL }%
  \providecommand \Eprint [0]{\href }%
  \providecommand \doibase [0]{http://dx.doi.org/}%
  \providecommand \selectlanguage [0]{\@gobble}%
  \providecommand \bibinfo  [0]{\@secondoftwo}%
  \providecommand \bibfield  [0]{\@secondoftwo}%
  \providecommand \translation [1]{[#1]}%
  \providecommand \BibitemOpen [0]{}%
  \providecommand \bibitemStop [0]{}%
  \providecommand \bibitemNoStop [0]{.\EOS\space}%
  \providecommand \EOS [0]{\spacefactor3000\relax}%
  \providecommand \BibitemShut  [1]{\csname bibitem#1\endcsname}%
  \let\auto@bib@innerbib\@empty
  \bibitem [{\citenamefont {Sakurai}\ and\ \citenamefont
    {Napolitano}(2010)}]{sakurai2010modern}%
    \BibitemOpen
    \bibfield  {author} {\bibinfo {author} {\bibfnamefont {J.~J.}\ \bibnamefont
    {Sakurai}}\ and\ \bibinfo {author} {\bibfnamefont {J.~J.}\ \bibnamefont
    {Napolitano}},\ }\href
    {https://www.amazon.com/Modern-Quantum-Mechanics-2nd-Sakurai/dp/0805382917?SubscriptionId=AKIAIOBINVZYXZQZ2U3A&tag=chimbori05-20&linkCode=xm2&camp=2025&creative=165953&creativeASIN=0805382917}
    {\emph {\bibinfo {title} {Modern Quantum Mechanics (2nd Edition)}}}\
    (\bibinfo  {publisher} {Pearson},\ \bibinfo {year} {2010})\BibitemShut
    {NoStop}%
  \bibitem [{\citenamefont {Siegman}(1986)}]{siegman1986lasers}%
    \BibitemOpen
    \bibfield  {author} {\bibinfo {author} {\bibfnamefont {A.~E.}\ \bibnamefont
    {Siegman}},\ }\href@noop {} {\emph {\bibinfo {title} {Lasers}}}\ (\bibinfo
    {publisher} {University Science Books},\ \bibinfo {address} {Sausalito, CA},\
    \bibinfo {year} {1986})\BibitemShut {NoStop}%
  \bibitem [{\citenamefont {Ince}(1925)}]{ince1925linear}%
    \BibitemOpen
    \bibfield  {author} {\bibinfo {author} {\bibfnamefont {E.~L.}\ \bibnamefont
    {Ince}},\ }\href {\doibase 10.1112/plms/s2-23.1.56} {\bibfield  {journal}
    {\bibinfo  {journal} {Proceedings of the London Mathematical Society}\
    }\textbf {\bibinfo {volume} {s2-23}},\ \bibinfo {pages} {56} (\bibinfo {year}
    {1925})}\BibitemShut {NoStop}%
  \bibitem [{\citenamefont {Arscott}(1964)}]{arscott1964periodic}%
    \BibitemOpen
    \bibfield  {author} {\bibinfo {author} {\bibfnamefont {F.~M.}\ \bibnamefont
    {Arscott}},\ }\href@noop {} {\emph {\bibinfo {title} {Periodic Differential
    Equations}}}\ (\bibinfo  {publisher} {Pergamon Press},\ \bibinfo {year}
    {1964})\BibitemShut {NoStop}%
  \bibitem [{\citenamefont {{Abbott}}\ \emph {et~al.}(2016)\citenamefont
    {{Abbott}} \emph {et~al.}}]{abbot2016observation}%
    \BibitemOpen
    \bibfield  {author} {\bibinfo {author} {\bibfnamefont {B.~P.}\ \bibnamefont
    {{Abbott}}} \emph {et~al.} (\bibinfo {collaboration} {LIGO Scientific
    Collaboration and Virgo Collaboration}),\ }\href {\doibase
    10.1103/PhysRevLett.116.061102} {\bibfield  {journal} {\bibinfo  {journal}
    {Phys. Rev. Lett.}\ }\textbf {\bibinfo {volume} {116}},\ \bibinfo {pages}
    {061102} (\bibinfo {year} {2016})}\BibitemShut {NoStop}%
  \bibitem [{\citenamefont {Tao}\ \emph {et~al.}(2020)\citenamefont {Tao},
    \citenamefont {Green},\ and\ \citenamefont {Fulda}}]{tao2020higher}%
    \BibitemOpen
    \bibfield  {author} {\bibinfo {author} {\bibfnamefont {L.}~\bibnamefont
    {Tao}}, \bibinfo {author} {\bibfnamefont {A.}~\bibnamefont {Green}}, \ and\
    \bibinfo {author} {\bibfnamefont {P.}~\bibnamefont {Fulda}},\ }\href
    {\doibase 10.1103/physrevd.102.122002} {\ \textbf {\bibinfo {volume} {102}},\
    \bibinfo {pages} {122002} (\bibinfo {year} {2020})}\BibitemShut {NoStop}%
  \bibitem [{\citenamefont {Boyer}\ \emph
    {et~al.}(1975{\natexlab{a}})\citenamefont {Boyer}, \citenamefont {Kalnins},\
    and\ \citenamefont {Miller}}]{boyer1975liea}%
    \BibitemOpen
    \bibfield  {author} {\bibinfo {author} {\bibfnamefont {C.~P.}\ \bibnamefont
    {Boyer}}, \bibinfo {author} {\bibfnamefont {E.~G.}\ \bibnamefont {Kalnins}},
    \ and\ \bibinfo {author} {\bibfnamefont {W.}~\bibnamefont {Miller}},\ }\href
    {\doibase 10.1063/1.522573} {\bibfield  {journal} {\bibinfo  {journal} {J.
    Math. Phys.}\ }\textbf {\bibinfo {volume} {16}},\ \bibinfo {pages} {499}
    (\bibinfo {year} {1975}{\natexlab{a}})}\BibitemShut {NoStop}%
  \bibitem [{\citenamefont {Boyer}\ \emph
    {et~al.}(1975{\natexlab{b}})\citenamefont {Boyer}, \citenamefont {Kalnins},\
    and\ \citenamefont {Miller}}]{boyer1975lie}%
    \BibitemOpen
    \bibfield  {author} {\bibinfo {author} {\bibfnamefont {C.~P.}\ \bibnamefont
    {Boyer}}, \bibinfo {author} {\bibfnamefont {E.~G.}\ \bibnamefont {Kalnins}},
    \ and\ \bibinfo {author} {\bibfnamefont {W.}~\bibnamefont {Miller}},\ }\href
    {\doibase 10.1063/1.522574} {\bibfield  {journal} {\bibinfo  {journal} {J.
    Math. Phys.}\ }\textbf {\bibinfo {volume} {16}},\ \bibinfo {pages} {512}
    (\bibinfo {year} {1975}{\natexlab{b}})}\BibitemShut {NoStop}%
  \bibitem [{\citenamefont {Bandres}\ and\ \citenamefont
    {Guti{\'{e}}rrez-Vega}(2004{\natexlab{a}})}]{bandres2004incegaussiana}%
    \BibitemOpen
    \bibfield  {author} {\bibinfo {author} {\bibfnamefont {M.~A.}\ \bibnamefont
    {Bandres}}\ and\ \bibinfo {author} {\bibfnamefont {J.~C.}\ \bibnamefont
    {Guti{\'{e}}rrez-Vega}},\ }\href {\doibase 10.1364/ol.29.000144} {\bibfield
    {journal} {\bibinfo  {journal} {Opt. Lett.}\ }\textbf {\bibinfo {volume}
    {29}},\ \bibinfo {pages} {144} (\bibinfo {year}
    {2004}{\natexlab{a}})}\BibitemShut {NoStop}%
  \bibitem [{\citenamefont {Bandres}\ and\ \citenamefont
    {Guti{\'{e}}rrez-Vega}(2004{\natexlab{b}})}]{bandres2004incegaussian}%
    \BibitemOpen
    \bibfield  {author} {\bibinfo {author} {\bibfnamefont {M.~A.}\ \bibnamefont
    {Bandres}}\ and\ \bibinfo {author} {\bibfnamefont {J.~C.}\ \bibnamefont
    {Guti{\'{e}}rrez-Vega}},\ }\href {\doibase 10.1364/josaa.21.000873}
    {\bibfield  {journal} {\bibinfo  {journal} {J. Opt. Soc. Am. A}\ }\textbf
    {\bibinfo {volume} {21}},\ \bibinfo {pages} {873} (\bibinfo {year}
    {2004}{\natexlab{b}})}\BibitemShut {NoStop}%
  \bibitem [{\citenamefont {Schwarz}\ \emph {et~al.}(2004)\citenamefont
    {Schwarz}, \citenamefont {Bandres},\ and\ \citenamefont
    {Guti{\'{e}}rrez-Vega}}]{schwarz2004observation}%
    \BibitemOpen
    \bibfield  {author} {\bibinfo {author} {\bibfnamefont {U.~T.}\ \bibnamefont
    {Schwarz}}, \bibinfo {author} {\bibfnamefont {M.~A.}\ \bibnamefont
    {Bandres}}, \ and\ \bibinfo {author} {\bibfnamefont {J.~C.}\ \bibnamefont
    {Guti{\'{e}}rrez-Vega}},\ }\href {\doibase 10.1364/ol.29.001870} {\bibfield
    {journal} {\bibinfo  {journal} {Opt. Lett.}\ }\textbf {\bibinfo {volume}
    {29}},\ \bibinfo {pages} {1870} (\bibinfo {year} {2004})}\BibitemShut
    {NoStop}%
  \bibitem [{\citenamefont {Krenn}\ \emph {et~al.}(2013)\citenamefont {Krenn},
    \citenamefont {Fickler}, \citenamefont {Huber}, \citenamefont {Lapkiewicz},
    \citenamefont {Plick}, \citenamefont {Ramelow},\ and\ \citenamefont
    {Zeilinger}}]{krenn2013entangled}%
    \BibitemOpen
    \bibfield  {author} {\bibinfo {author} {\bibfnamefont {M.}~\bibnamefont
    {Krenn}}, \bibinfo {author} {\bibfnamefont {R.}~\bibnamefont {Fickler}},
    \bibinfo {author} {\bibfnamefont {M.}~\bibnamefont {Huber}}, \bibinfo
    {author} {\bibfnamefont {R.}~\bibnamefont {Lapkiewicz}}, \bibinfo {author}
    {\bibfnamefont {W.}~\bibnamefont {Plick}}, \bibinfo {author} {\bibfnamefont
    {S.}~\bibnamefont {Ramelow}}, \ and\ \bibinfo {author} {\bibfnamefont
    {A.}~\bibnamefont {Zeilinger}},\ }\href {\doibase 10.1103/physreva.87.012326}
    {\bibfield  {journal} {\bibinfo  {journal} {Phys. Rev. A}\ }\textbf {\bibinfo
    {volume} {87}},\ \bibinfo {pages} {012326} (\bibinfo {year}
    {2013})}\BibitemShut {NoStop}%
  \bibitem [{\citenamefont {Plick}\ \emph {et~al.}(2013)\citenamefont {Plick},
    \citenamefont {Krenn}, \citenamefont {Fickler}, \citenamefont {Ramelow},\
    and\ \citenamefont {Zeilinger}}]{plick2013quantum}%
    \BibitemOpen
    \bibfield  {author} {\bibinfo {author} {\bibfnamefont {W.~N.}\ \bibnamefont
    {Plick}}, \bibinfo {author} {\bibfnamefont {M.}~\bibnamefont {Krenn}},
    \bibinfo {author} {\bibfnamefont {R.}~\bibnamefont {Fickler}}, \bibinfo
    {author} {\bibfnamefont {S.}~\bibnamefont {Ramelow}}, \ and\ \bibinfo
    {author} {\bibfnamefont {A.}~\bibnamefont {Zeilinger}},\ }\href {\doibase
    10.1103/physreva.87.033806} {\bibfield  {journal} {\bibinfo  {journal} {Phys.
    Rev. A}\ }\textbf {\bibinfo {volume} {87}},\ \bibinfo {pages} {033806}
    (\bibinfo {year} {2013})}\BibitemShut {NoStop}%
  \bibitem [{\citenamefont {Woerdemann}\ \emph {et~al.}(2011)\citenamefont
    {Woerdemann}, \citenamefont {Alpmann},\ and\ \citenamefont
    {Denz}}]{woerdemann2011optical}%
    \BibitemOpen
    \bibfield  {author} {\bibinfo {author} {\bibfnamefont {M.}~\bibnamefont
    {Woerdemann}}, \bibinfo {author} {\bibfnamefont {C.}~\bibnamefont {Alpmann}},
    \ and\ \bibinfo {author} {\bibfnamefont {C.}~\bibnamefont {Denz}},\ }\href
    {\doibase 10.1063/1.3561770} {\bibfield  {journal} {\bibinfo  {journal}
    {Appl. Phys. Lett.}\ }\textbf {\bibinfo {volume} {98}},\ \bibinfo {pages}
    {111101} (\bibinfo {year} {2011})}\BibitemShut {NoStop}%
  \bibitem [{\citenamefont {Milburn}\ \emph {et~al.}(1997)\citenamefont
    {Milburn}, \citenamefont {Corney}, \citenamefont {Wright},\ and\
    \citenamefont {Walls}}]{milburn1997quantum}%
    \BibitemOpen
    \bibfield  {author} {\bibinfo {author} {\bibfnamefont {G.~J.}\ \bibnamefont
    {Milburn}}, \bibinfo {author} {\bibfnamefont {J.}~\bibnamefont {Corney}},
    \bibinfo {author} {\bibfnamefont {E.~M.}\ \bibnamefont {Wright}}, \ and\
    \bibinfo {author} {\bibfnamefont {D.~F.}\ \bibnamefont {Walls}},\ }\href
    {\doibase 10.1103/physreva.55.4318} {\bibfield  {journal} {\bibinfo
    {journal} {Phys. Rev. A}\ }\textbf {\bibinfo {volume} {55}},\ \bibinfo
    {pages} {4318} (\bibinfo {year} {1997})}\BibitemShut {NoStop}%
  \bibitem [{\citenamefont {Smerzi}\ \emph {et~al.}(1997)\citenamefont {Smerzi},
    \citenamefont {Fantoni}, \citenamefont {Giovanazzi},\ and\ \citenamefont
    {Shenoy}}]{smerzi1997quantum}%
    \BibitemOpen
    \bibfield  {author} {\bibinfo {author} {\bibfnamefont {A.}~\bibnamefont
    {Smerzi}}, \bibinfo {author} {\bibfnamefont {S.}~\bibnamefont {Fantoni}},
    \bibinfo {author} {\bibfnamefont {S.}~\bibnamefont {Giovanazzi}}, \ and\
    \bibinfo {author} {\bibfnamefont {S.~R.}\ \bibnamefont {Shenoy}},\ }\href
    {\doibase 10.1103/physrevlett.79.4950} {\bibfield  {journal} {\bibinfo
    {journal} {Phys. Rev. Lett.}\ }\textbf {\bibinfo {volume} {79}},\ \bibinfo
    {pages} {4950} (\bibinfo {year} {1997})}\BibitemShut {NoStop}%
  \bibitem [{\citenamefont {Raghavan}\ \emph {et~al.}(1999)\citenamefont
    {Raghavan}, \citenamefont {Smerzi}, \citenamefont {Fantoni},\ and\
    \citenamefont {Shenoy}}]{raghavan1999coherent}%
    \BibitemOpen
    \bibfield  {author} {\bibinfo {author} {\bibfnamefont {S.}~\bibnamefont
    {Raghavan}}, \bibinfo {author} {\bibfnamefont {A.}~\bibnamefont {Smerzi}},
    \bibinfo {author} {\bibfnamefont {S.}~\bibnamefont {Fantoni}}, \ and\
    \bibinfo {author} {\bibfnamefont {S.~R.}\ \bibnamefont {Shenoy}},\ }\href
    {\doibase 10.1103/physreva.59.620} {\bibfield  {journal} {\bibinfo  {journal}
    {Phys. Rev. A}\ }\textbf {\bibinfo {volume} {59}},\ \bibinfo {pages} {620}
    (\bibinfo {year} {1999})}\BibitemShut {NoStop}%
  \bibitem [{\citenamefont {Anglin}\ \emph {et~al.}(2001)\citenamefont {Anglin},
    \citenamefont {Drummond},\ and\ \citenamefont {Smerzi}}]{anglin2001exact}%
    \BibitemOpen
    \bibfield  {author} {\bibinfo {author} {\bibfnamefont {J.~R.}\ \bibnamefont
    {Anglin}}, \bibinfo {author} {\bibfnamefont {P.}~\bibnamefont {Drummond}}, \
    and\ \bibinfo {author} {\bibfnamefont {A.}~\bibnamefont {Smerzi}},\ }\href
    {\doibase 10.1103/physreva.64.063605} {\bibfield  {journal} {\bibinfo
    {journal} {Phys. Rev. A}\ }\textbf {\bibinfo {volume} {64}},\ \bibinfo
    {pages} {063605} (\bibinfo {year} {2001})}\BibitemShut {NoStop}%
  \bibitem [{\citenamefont {Leggett}(2001)}]{leggett2001bose}%
    \BibitemOpen
    \bibfield  {author} {\bibinfo {author} {\bibfnamefont {A.~J.}\ \bibnamefont
    {Leggett}},\ }\href {\doibase 10.1103/revmodphys.73.307} {\bibfield
    {journal} {\bibinfo  {journal} {Rev Mod Phys}\ }\textbf {\bibinfo {volume}
    {73}},\ \bibinfo {pages} {307} (\bibinfo {year} {2001})}\BibitemShut
    {NoStop}%
  \bibitem [{\citenamefont {Gati}\ and\ \citenamefont
    {Oberthaler}(2007)}]{gati2007bosonic}%
    \BibitemOpen
    \bibfield  {author} {\bibinfo {author} {\bibfnamefont {R.}~\bibnamefont
    {Gati}}\ and\ \bibinfo {author} {\bibfnamefont {M.~K.}\ \bibnamefont
    {Oberthaler}},\ }\href {\doibase 10.1088/0953-4075/40/10/r01} {\bibfield
    {journal} {\bibinfo  {journal} {J. Phys. B: At., Mol. Opt. Phys.}\ }\textbf
    {\bibinfo {volume} {40}},\ \bibinfo {pages} {R61} (\bibinfo {year}
    {2007})}\BibitemShut {NoStop}%
  \bibitem [{\citenamefont {Graefe}\ and\ \citenamefont
    {Korsch}(2007)}]{graefe2007semiclassical}%
    \BibitemOpen
    \bibfield  {author} {\bibinfo {author} {\bibfnamefont {E.~M.}\ \bibnamefont
    {Graefe}}\ and\ \bibinfo {author} {\bibfnamefont {H.~J.}\ \bibnamefont
    {Korsch}},\ }\href {\doibase 10.1103/PhysRevA.76.032116} {\bibfield
    {journal} {\bibinfo  {journal} {Phys. Rev. A}\ }\textbf {\bibinfo {volume}
    {76}},\ \bibinfo {pages} {032116} (\bibinfo {year} {2007})}\BibitemShut
    {NoStop}%
  \bibitem [{\citenamefont {Sakmann}\ \emph {et~al.}(2009)\citenamefont
    {Sakmann}, \citenamefont {Streltsov}, \citenamefont {Alon},\ and\
    \citenamefont {Cederbaum}}]{sakmann2009exact}%
    \BibitemOpen
    \bibfield  {author} {\bibinfo {author} {\bibfnamefont {K.}~\bibnamefont
    {Sakmann}}, \bibinfo {author} {\bibfnamefont {A.~I.}\ \bibnamefont
    {Streltsov}}, \bibinfo {author} {\bibfnamefont {O.~E.}\ \bibnamefont {Alon}},
    \ and\ \bibinfo {author} {\bibfnamefont {L.~S.}\ \bibnamefont {Cederbaum}},\
    }\href {\doibase 10.1103/PhysRevLett.103.220601} {\bibfield  {journal}
    {\bibinfo  {journal} {Phys. Rev. Lett.}\ }\textbf {\bibinfo {volume} {103}},\
    \bibinfo {pages} {220601} (\bibinfo {year} {2009})}\BibitemShut {NoStop}%
  \bibitem [{\citenamefont {Nissen}\ and\ \citenamefont
    {Keeling}(2010)}]{nissen2010wentzel}%
    \BibitemOpen
    \bibfield  {author} {\bibinfo {author} {\bibfnamefont {F.}~\bibnamefont
    {Nissen}}\ and\ \bibinfo {author} {\bibfnamefont {J.}~\bibnamefont
    {Keeling}},\ }\href {\doibase 10.1103/PhysRevA.81.063628} {\bibfield
    {journal} {\bibinfo  {journal} {Phys. Rev. A}\ }\textbf {\bibinfo {volume}
    {81}},\ \bibinfo {pages} {063628} (\bibinfo {year} {2010})}\BibitemShut
    {NoStop}%
  \bibitem [{\citenamefont {Chuchem}\ \emph {et~al.}(2010)\citenamefont
    {Chuchem}, \citenamefont {Smith-Mannschott}, \citenamefont {Hiller},
    \citenamefont {Kottos}, \citenamefont {Vardi},\ and\ \citenamefont
    {Cohen}}]{chuchem2010quantum}%
    \BibitemOpen
    \bibfield  {author} {\bibinfo {author} {\bibfnamefont {M.}~\bibnamefont
    {Chuchem}}, \bibinfo {author} {\bibfnamefont {K.}~\bibnamefont
    {Smith-Mannschott}}, \bibinfo {author} {\bibfnamefont {M.}~\bibnamefont
    {Hiller}}, \bibinfo {author} {\bibfnamefont {T.}~\bibnamefont {Kottos}},
    \bibinfo {author} {\bibfnamefont {A.}~\bibnamefont {Vardi}}, \ and\ \bibinfo
    {author} {\bibfnamefont {D.}~\bibnamefont {Cohen}},\ }\href {\doibase
    10.1103/physreva.82.053617} {\bibfield  {journal} {\bibinfo  {journal} {Phys.
    Rev. A}\ }\textbf {\bibinfo {volume} {82}},\ \bibinfo {pages} {053617}
    (\bibinfo {year} {2010})}\BibitemShut {NoStop}%
  \bibitem [{\citenamefont {O'Dell}(2012)}]{odell2012quantum}%
    \BibitemOpen
    \bibfield  {author} {\bibinfo {author} {\bibfnamefont {D.~H.~J.}\
    \bibnamefont {O'Dell}},\ }\href {\doibase 10.1103/physrevlett.109.150406}
    {\bibfield  {journal} {\bibinfo  {journal} {Phys. Rev. Lett.}\ }\textbf
    {\bibinfo {volume} {109}},\ \bibinfo {pages} {150406} (\bibinfo {year}
    {2012})}\BibitemShut {NoStop}%
  \bibitem [{\citenamefont {Graefe}\ \emph {et~al.}(2014)\citenamefont {Graefe},
    \citenamefont {Korsch},\ and\ \citenamefont
    {Strzys}}]{graefe2014bosehubbard}%
    \BibitemOpen
    \bibfield  {author} {\bibinfo {author} {\bibfnamefont {E.-M.}\ \bibnamefont
    {Graefe}}, \bibinfo {author} {\bibfnamefont {H.~J.}\ \bibnamefont {Korsch}},
    \ and\ \bibinfo {author} {\bibfnamefont {M.~P.}\ \bibnamefont {Strzys}},\
    }\href {\doibase 10.1088/1751-8113/47/8/085304} {\bibfield  {journal}
    {\bibinfo  {journal} {J. Phys. A: Math. Theor.}\ }\textbf {\bibinfo {volume}
    {47}},\ \bibinfo {pages} {085304} (\bibinfo {year} {2014})}\BibitemShut
    {NoStop}%
  \bibitem [{\citenamefont {Sukhatme}\ \emph {et~al.}(2001)\citenamefont
    {Sukhatme}, \citenamefont {Mukharsky}, \citenamefont {Chui},\ and\
    \citenamefont {Pearson}}]{sukhatme2001observation}%
    \BibitemOpen
    \bibfield  {author} {\bibinfo {author} {\bibfnamefont {K.}~\bibnamefont
    {Sukhatme}}, \bibinfo {author} {\bibfnamefont {Y.}~\bibnamefont {Mukharsky}},
    \bibinfo {author} {\bibfnamefont {T.}~\bibnamefont {Chui}}, \ and\ \bibinfo
    {author} {\bibfnamefont {D.}~\bibnamefont {Pearson}},\ }\href {\doibase
    10.1038/35077024} {\bibfield  {journal} {\bibinfo  {journal} {Nature}\
    }\textbf {\bibinfo {volume} {411}},\ \bibinfo {pages} {280} (\bibinfo {year}
    {2001})}\BibitemShut {NoStop}%
  \bibitem [{\citenamefont {Backhaus}\ \emph {et~al.}(1998)\citenamefont
    {Backhaus}, \citenamefont {Pereverzev}, \citenamefont {Simmonds},
    \citenamefont {Loshak}, \citenamefont {Davis},\ and\ \citenamefont
    {Packard}}]{backhaus1998discovery}%
    \BibitemOpen
    \bibfield  {author} {\bibinfo {author} {\bibfnamefont {S.}~\bibnamefont
    {Backhaus}}, \bibinfo {author} {\bibfnamefont {S.}~\bibnamefont
    {Pereverzev}}, \bibinfo {author} {\bibfnamefont {R.~W.}\ \bibnamefont
    {Simmonds}}, \bibinfo {author} {\bibfnamefont {A.}~\bibnamefont {Loshak}},
    \bibinfo {author} {\bibfnamefont {J.~C.}\ \bibnamefont {Davis}}, \ and\
    \bibinfo {author} {\bibfnamefont {R.~E.}\ \bibnamefont {Packard}},\ }\href
    {\doibase 10.1038/33629} {\bibfield  {journal} {\bibinfo  {journal} {Nature}\
    }\textbf {\bibinfo {volume} {392}},\ \bibinfo {pages} {687} (\bibinfo {year}
    {1998})}\BibitemShut {NoStop}%
  \bibitem [{\citenamefont {Cataliotti}\ \emph {et~al.}(2001)\citenamefont
    {Cataliotti}, \citenamefont {Burger}, \citenamefont {Fort}, \citenamefont
    {Maddaloni}, \citenamefont {Minardi}, \citenamefont {Trombettoni},
    \citenamefont {Smerzi},\ and\ \citenamefont
    {Inguscio}}]{cataliotti2001josephson}%
    \BibitemOpen
    \bibfield  {author} {\bibinfo {author} {\bibfnamefont {F.~S.}\ \bibnamefont
    {Cataliotti}}, \bibinfo {author} {\bibfnamefont {S.}~\bibnamefont {Burger}},
    \bibinfo {author} {\bibfnamefont {C.}~\bibnamefont {Fort}}, \bibinfo {author}
    {\bibfnamefont {P.}~\bibnamefont {Maddaloni}}, \bibinfo {author}
    {\bibfnamefont {F.}~\bibnamefont {Minardi}}, \bibinfo {author} {\bibfnamefont
    {A.}~\bibnamefont {Trombettoni}}, \bibinfo {author} {\bibfnamefont
    {A.}~\bibnamefont {Smerzi}}, \ and\ \bibinfo {author} {\bibfnamefont
    {M.}~\bibnamefont {Inguscio}},\ }\href {\doibase 10.1126/science.1062612}
    {\bibfield  {journal} {\bibinfo  {journal} {Science}\ }\textbf {\bibinfo
    {volume} {293}},\ \bibinfo {pages} {843} (\bibinfo {year}
    {2001})}\BibitemShut {NoStop}%
  \bibitem [{\citenamefont {Albiez}\ \emph {et~al.}(2005)\citenamefont {Albiez},
    \citenamefont {Gati}, \citenamefont {Fölling}, \citenamefont {Hunsmann},
    \citenamefont {Cristiani},\ and\ \citenamefont
    {Oberthaler}}]{albiez2005direct}%
    \BibitemOpen
    \bibfield  {author} {\bibinfo {author} {\bibfnamefont {M.}~\bibnamefont
    {Albiez}}, \bibinfo {author} {\bibfnamefont {R.}~\bibnamefont {Gati}},
    \bibinfo {author} {\bibfnamefont {J.}~\bibnamefont {Fölling}}, \bibinfo
    {author} {\bibfnamefont {S.}~\bibnamefont {Hunsmann}}, \bibinfo {author}
    {\bibfnamefont {M.}~\bibnamefont {Cristiani}}, \ and\ \bibinfo {author}
    {\bibfnamefont {M.~K.}\ \bibnamefont {Oberthaler}},\ }\href {\doibase
    10.1103/physrevlett.95.010402} {\bibfield  {journal} {\bibinfo  {journal}
    {Phys. Rev. Lett.}\ }\textbf {\bibinfo {volume} {95}},\ \bibinfo {pages}
    {010402} (\bibinfo {year} {2005})}\BibitemShut {NoStop}%
  \bibitem [{\citenamefont {Schumm}\ \emph {et~al.}(2005)\citenamefont {Schumm},
    \citenamefont {Hofferberth}, \citenamefont {Andersson}, \citenamefont
    {Wildermuth}, \citenamefont {Groth}, \citenamefont {Bar-Joseph},
    \citenamefont {Schmiedmayer},\ and\ \citenamefont
    {Krüger}}]{schumm2005matter}%
    \BibitemOpen
    \bibfield  {author} {\bibinfo {author} {\bibfnamefont {T.}~\bibnamefont
    {Schumm}}, \bibinfo {author} {\bibfnamefont {S.}~\bibnamefont {Hofferberth}},
    \bibinfo {author} {\bibfnamefont {L.~M.}\ \bibnamefont {Andersson}}, \bibinfo
    {author} {\bibfnamefont {S.}~\bibnamefont {Wildermuth}}, \bibinfo {author}
    {\bibfnamefont {S.}~\bibnamefont {Groth}}, \bibinfo {author} {\bibfnamefont
    {I.}~\bibnamefont {Bar-Joseph}}, \bibinfo {author} {\bibfnamefont
    {J.}~\bibnamefont {Schmiedmayer}}, \ and\ \bibinfo {author} {\bibfnamefont
    {P.}~\bibnamefont {Krüger}},\ }\href {\doibase 10.1038/nphys125} {\bibfield
    {journal} {\bibinfo  {journal} {Nat. Phys.}\ }\textbf {\bibinfo {volume}
    {1}},\ \bibinfo {pages} {57} (\bibinfo {year} {2005})}\BibitemShut {NoStop}%
  \bibitem [{\citenamefont {Levy}\ \emph {et~al.}(2007)\citenamefont {Levy},
    \citenamefont {Lahoud}, \citenamefont {Shomroni},\ and\ \citenamefont
    {Steinhauer}}]{levy2007a.c.}%
    \BibitemOpen
    \bibfield  {author} {\bibinfo {author} {\bibfnamefont {S.}~\bibnamefont
    {Levy}}, \bibinfo {author} {\bibfnamefont {E.}~\bibnamefont {Lahoud}},
    \bibinfo {author} {\bibfnamefont {I.}~\bibnamefont {Shomroni}}, \ and\
    \bibinfo {author} {\bibfnamefont {J.}~\bibnamefont {Steinhauer}},\ }\href
    {\doibase 10.1038/nature06186} {\bibfield  {journal} {\bibinfo  {journal}
    {Nature}\ }\textbf {\bibinfo {volume} {449}},\ \bibinfo {pages} {579}
    (\bibinfo {year} {2007})}\BibitemShut {NoStop}%
  \bibitem [{\citenamefont {LeBlanc}\ \emph {et~al.}(2011)\citenamefont
    {LeBlanc}, \citenamefont {Bardon}, \citenamefont {McKeever}, \citenamefont
    {Extavour}, \citenamefont {Jervis}, \citenamefont {Thywissen}, \citenamefont
    {Piazza},\ and\ \citenamefont {Smerzi}}]{leblanc2011dynamics}%
    \BibitemOpen
    \bibfield  {author} {\bibinfo {author} {\bibfnamefont {L.~J.}\ \bibnamefont
    {LeBlanc}}, \bibinfo {author} {\bibfnamefont {A.~B.}\ \bibnamefont {Bardon}},
    \bibinfo {author} {\bibfnamefont {J.}~\bibnamefont {McKeever}}, \bibinfo
    {author} {\bibfnamefont {M.~H.~T.}\ \bibnamefont {Extavour}}, \bibinfo
    {author} {\bibfnamefont {D.}~\bibnamefont {Jervis}}, \bibinfo {author}
    {\bibfnamefont {J.~H.}\ \bibnamefont {Thywissen}}, \bibinfo {author}
    {\bibfnamefont {F.}~\bibnamefont {Piazza}}, \ and\ \bibinfo {author}
    {\bibfnamefont {A.}~\bibnamefont {Smerzi}},\ }\href {\doibase
    10.1103/physrevlett.106.025302} {\bibfield  {journal} {\bibinfo  {journal}
    {Phys. Rev. Lett.}\ }\textbf {\bibinfo {volume} {106}},\ \bibinfo {pages}
    {025302} (\bibinfo {year} {2011})}\BibitemShut {NoStop}%
  \bibitem [{\citenamefont {Gerving}\ \emph {et~al.}(2012)\citenamefont
    {Gerving}, \citenamefont {Hoang}, \citenamefont {Land}, \citenamefont
    {Anquez}, \citenamefont {Hamley},\ and\ \citenamefont
    {Chapman}}]{gerving2012non}%
    \BibitemOpen
    \bibfield  {author} {\bibinfo {author} {\bibfnamefont {C.}~\bibnamefont
    {Gerving}}, \bibinfo {author} {\bibfnamefont {T.}~\bibnamefont {Hoang}},
    \bibinfo {author} {\bibfnamefont {B.}~\bibnamefont {Land}}, \bibinfo {author}
    {\bibfnamefont {M.}~\bibnamefont {Anquez}}, \bibinfo {author} {\bibfnamefont
    {C.}~\bibnamefont {Hamley}}, \ and\ \bibinfo {author} {\bibfnamefont
    {M.}~\bibnamefont {Chapman}},\ }\href {\doibase 10.1038/ncomms2179}
    {\bibfield  {journal} {\bibinfo  {journal} {Nat Commun}\ }\textbf {\bibinfo
    {volume} {3}} (\bibinfo {year} {2012}),\ 10.1038/ncomms2179}\BibitemShut
    {NoStop}%
  \bibitem [{\citenamefont {Trenkwalder}\ \emph {et~al.}(2016)\citenamefont
    {Trenkwalder}, \citenamefont {Spagnolli}, \citenamefont {Semeghini},
    \citenamefont {Coop}, \citenamefont {Landini}, \citenamefont {Castilho},
    \citenamefont {Pezz{\`{e}}}, \citenamefont {Modugno}, \citenamefont
    {Inguscio}, \citenamefont {Smerzi},\ and\ \citenamefont
    {Fattori}}]{trenkwalder2016quantum}%
    \BibitemOpen
    \bibfield  {author} {\bibinfo {author} {\bibfnamefont {A.}~\bibnamefont
    {Trenkwalder}}, \bibinfo {author} {\bibfnamefont {G.}~\bibnamefont
    {Spagnolli}}, \bibinfo {author} {\bibfnamefont {G.}~\bibnamefont
    {Semeghini}}, \bibinfo {author} {\bibfnamefont {S.}~\bibnamefont {Coop}},
    \bibinfo {author} {\bibfnamefont {M.}~\bibnamefont {Landini}}, \bibinfo
    {author} {\bibfnamefont {P.}~\bibnamefont {Castilho}}, \bibinfo {author}
    {\bibfnamefont {L.}~\bibnamefont {Pezz{\`{e}}}}, \bibinfo {author}
    {\bibfnamefont {G.}~\bibnamefont {Modugno}}, \bibinfo {author} {\bibfnamefont
    {M.}~\bibnamefont {Inguscio}}, \bibinfo {author} {\bibfnamefont
    {A.}~\bibnamefont {Smerzi}}, \ and\ \bibinfo {author} {\bibfnamefont
    {M.}~\bibnamefont {Fattori}},\ }\href {\doibase 10.1038/nphys3743} {\bibfield
     {journal} {\bibinfo  {journal} {Nat. Phys.}\ }\textbf {\bibinfo {volume}
    {12}},\ \bibinfo {pages} {826} (\bibinfo {year} {2016})}\BibitemShut
    {NoStop}%
  \bibitem [{\citenamefont {Lagoudakis}\ \emph {et~al.}(2010)\citenamefont
    {Lagoudakis}, \citenamefont {Pietka}, \citenamefont {Wouters}, \citenamefont
    {Andr{\'{e}}},\ and\ \citenamefont
    {Deveaud-Pl{\'{e}}dran}}]{lagoudakis2010coherent}%
    \BibitemOpen
    \bibfield  {author} {\bibinfo {author} {\bibfnamefont {K.~G.}\ \bibnamefont
    {Lagoudakis}}, \bibinfo {author} {\bibfnamefont {B.}~\bibnamefont {Pietka}},
    \bibinfo {author} {\bibfnamefont {M.}~\bibnamefont {Wouters}}, \bibinfo
    {author} {\bibfnamefont {R.}~\bibnamefont {Andr{\'{e}}}}, \ and\ \bibinfo
    {author} {\bibfnamefont {B.}~\bibnamefont {Deveaud-Pl{\'{e}}dran}},\ }\href
    {\doibase 10.1103/physrevlett.105.120403} {\bibfield  {journal} {\bibinfo
    {journal} {Phys Rev Lett}\ }\textbf {\bibinfo {volume} {105}},\ \bibinfo
    {pages} {120403} (\bibinfo {year} {2010})}\BibitemShut {NoStop}%
  \bibitem [{\citenamefont {Abbarchi}\ \emph {et~al.}(2013)\citenamefont
    {Abbarchi}, \citenamefont {Amo}, \citenamefont {Sala}, \citenamefont
    {Solnyshkov}, \citenamefont {Flayac}, \citenamefont {Ferrier}, \citenamefont
    {Sagnes}, \citenamefont {Galopin}, \citenamefont {Lema{\^{\i}}tre},
    \citenamefont {Malpuech},\ and\ \citenamefont
    {Bloch}}]{abbarchi2013macroscopic}%
    \BibitemOpen
    \bibfield  {author} {\bibinfo {author} {\bibfnamefont {M.}~\bibnamefont
    {Abbarchi}}, \bibinfo {author} {\bibfnamefont {A.}~\bibnamefont {Amo}},
    \bibinfo {author} {\bibfnamefont {V.~G.}\ \bibnamefont {Sala}}, \bibinfo
    {author} {\bibfnamefont {D.~D.}\ \bibnamefont {Solnyshkov}}, \bibinfo
    {author} {\bibfnamefont {H.}~\bibnamefont {Flayac}}, \bibinfo {author}
    {\bibfnamefont {L.}~\bibnamefont {Ferrier}}, \bibinfo {author} {\bibfnamefont
    {I.}~\bibnamefont {Sagnes}}, \bibinfo {author} {\bibfnamefont
    {E.}~\bibnamefont {Galopin}}, \bibinfo {author} {\bibfnamefont
    {A.}~\bibnamefont {Lema{\^{\i}}tre}}, \bibinfo {author} {\bibfnamefont
    {G.}~\bibnamefont {Malpuech}}, \ and\ \bibinfo {author} {\bibfnamefont
    {J.}~\bibnamefont {Bloch}},\ }\href {\doibase 10.1038/nphys2609} {\bibfield
    {journal} {\bibinfo  {journal} {Nat Phys}\ }\textbf {\bibinfo {volume} {9}},\
    \bibinfo {pages} {275} (\bibinfo {year} {2013})}\BibitemShut {NoStop}%
  \bibitem [{\citenamefont {Guti{\'{e}}rrez-Cuevas}\ \emph
    {et~al.}(2020)\citenamefont {Guti{\'{e}}rrez-Cuevas}, \citenamefont {Wadood},
    \citenamefont {Vamivakas},\ and\ \citenamefont
    {Alonso}}]{gutierrez-cuevas2020modal}%
    \BibitemOpen
    \bibfield  {author} {\bibinfo {author} {\bibfnamefont {R.}~\bibnamefont
    {Guti{\'{e}}rrez-Cuevas}}, \bibinfo {author} {\bibfnamefont {S.~A.}\
    \bibnamefont {Wadood}}, \bibinfo {author} {\bibfnamefont {A.~N.}\
    \bibnamefont {Vamivakas}}, \ and\ \bibinfo {author} {\bibfnamefont {M.~A.}\
    \bibnamefont {Alonso}},\ }\href {\doibase 10.1103/physrevlett.125.123903}
    {\bibfield  {journal} {\bibinfo  {journal} {Phys. Rev. Lett.}\ }\textbf
    {\bibinfo {volume} {125}},\ \bibinfo {pages} {123903} (\bibinfo {year}
    {2020})}\BibitemShut {NoStop}%
  \bibitem [{\citenamefont {Dennis}\ and\ \citenamefont
    {Alonso}(2017)}]{dennis2017swings}%
    \BibitemOpen
    \bibfield  {author} {\bibinfo {author} {\bibfnamefont {M.~R.}\ \bibnamefont
    {Dennis}}\ and\ \bibinfo {author} {\bibfnamefont {M.~A.}\ \bibnamefont
    {Alonso}},\ }\href {\doibase 10.1098/rsta.2015.0441} {\bibfield  {journal}
    {\bibinfo  {journal} {Phil. Trans. R. Soc. A}\ }\textbf {\bibinfo {volume}
    {375}},\ \bibinfo {pages} {20150441} (\bibinfo {year} {2017})}\BibitemShut
    {NoStop}%
  \bibitem [{\citenamefont {Guti{\'{e}}rrez-Cuevas}\ \emph
    {et~al.}(2019)\citenamefont {Guti{\'{e}}rrez-Cuevas}, \citenamefont
    {Dennis},\ and\ \citenamefont {Alonso}}]{gutierrez-cuevas2019generalized}%
    \BibitemOpen
    \bibfield  {author} {\bibinfo {author} {\bibfnamefont {R.}~\bibnamefont
    {Guti{\'{e}}rrez-Cuevas}}, \bibinfo {author} {\bibfnamefont {M.~R.}\
    \bibnamefont {Dennis}}, \ and\ \bibinfo {author} {\bibfnamefont {M.~A.}\
    \bibnamefont {Alonso}},\ }\href {\doibase 10.1088/2040-8986/ab2c52}
    {\bibfield  {journal} {\bibinfo  {journal} {J. Opt.}\ }\textbf {\bibinfo
    {volume} {21}},\ \bibinfo {pages} {084001} (\bibinfo {year}
    {2019})}\BibitemShut {NoStop}%
  \bibitem [{\citenamefont {Dennis}\ and\ \citenamefont
    {Alonso}(2019)}]{dennis2019gaussian}%
    \BibitemOpen
    \bibfield  {author} {\bibinfo {author} {\bibfnamefont {M.~R.}\ \bibnamefont
    {Dennis}}\ and\ \bibinfo {author} {\bibfnamefont {M.~A.}\ \bibnamefont
    {Alonso}},\ }\href {\doibase 10.1088/2515-7647/ab011d} {\bibfield  {journal}
    {\bibinfo  {journal} {J. Phys. Photonics}\ }\textbf {\bibinfo {volume} {1}},\
    \bibinfo {pages} {025003} (\bibinfo {year} {2019})}\BibitemShut {NoStop}%
  \bibitem [{\citenamefont {Guti{\'{e}}rrez-Cuevas}\ and\ \citenamefont
    {Alonso}(2020)}]{gutierrez-cuevas2020platonic}%
    \BibitemOpen
    \bibfield  {author} {\bibinfo {author} {\bibfnamefont {R.}~\bibnamefont
    {Guti{\'{e}}rrez-Cuevas}}\ and\ \bibinfo {author} {\bibfnamefont {M.~A.}\
    \bibnamefont {Alonso}},\ }\href {\doibase 10.1364/ol.405988} {\bibfield
    {journal} {\bibinfo  {journal} {Opt Lett}\ }\textbf {\bibinfo {volume}
    {45}},\ \bibinfo {pages} {6759} (\bibinfo {year} {2020})}\BibitemShut
    {NoStop}%
  \bibitem [{SM()}]{SM}%
    \BibitemOpen
    \href@noop {} {\bibinfo  {journal} {See Supplemental Material at [URL will be
    inserted by publisher] for details about the Ince-Gauss modes, the derivation
    of the operator for the aberrated cavity and background information on the
    Bose-Hubbard dimer, which include Refs. [44-62]}\ }\BibitemShut {NoStop}%
  \bibitem [{\citenamefont {Alonso}(2011)}]{alonso2011wigner}%
    \BibitemOpen
  \bibfield  {journal} {  }\bibfield  {author} {\bibinfo {author} {\bibfnamefont
    {M.~A.}\ \bibnamefont {Alonso}},\ }\href {\doibase 10.1364/aop.3.000272}
    {\bibfield  {journal} {\bibinfo  {journal} {Adv. Opt. Photonics}\ }\textbf
    {\bibinfo {volume} {3}},\ \bibinfo {pages} {272} (\bibinfo {year}
    {2011})}\BibitemShut {NoStop}%
  \bibitem [{\citenamefont {Healy}\ \emph {et~al.}(2015)\citenamefont {Healy},
    \citenamefont {Kutay}, \citenamefont {Ozaktas},\ and\ \citenamefont
    {Sheridan}}]{healy2015linear}%
    \BibitemOpen
    \bibfield  {author} {\bibinfo {author} {\bibfnamefont {J.~J.}\ \bibnamefont
    {Healy}}, \bibinfo {author} {\bibfnamefont {M.~A.}\ \bibnamefont {Kutay}},
    \bibinfo {author} {\bibfnamefont {H.~M.}\ \bibnamefont {Ozaktas}}, \ and\
    \bibinfo {author} {\bibfnamefont {J.~T.}\ \bibnamefont {Sheridan}},\
    }\href@noop {} {\emph {\bibinfo {title} {Linear canonical transforms: Theory
    and applications}}},\ Vol.\ \bibinfo {volume} {198}\ (\bibinfo  {publisher}
    {Springer},\ \bibinfo {year} {2015})\BibitemShut {NoStop}%
  \bibitem [{\citenamefont {Lipkin}\ \emph {et~al.}(1965)\citenamefont {Lipkin},
    \citenamefont {Meshkov},\ and\ \citenamefont {Glick}}]{lipkin1965validity}%
    \BibitemOpen
    \bibfield  {author} {\bibinfo {author} {\bibfnamefont {H.}~\bibnamefont
    {Lipkin}}, \bibinfo {author} {\bibfnamefont {N.}~\bibnamefont {Meshkov}}, \
    and\ \bibinfo {author} {\bibfnamefont {A.}~\bibnamefont {Glick}},\ }\href
    {\doibase 10.1016/0029-5582(65)90862-x} {\bibfield  {journal} {\bibinfo
    {journal} {Nuclear Physics}\ }\textbf {\bibinfo {volume} {62}},\ \bibinfo
    {pages} {188} (\bibinfo {year} {1965})}\BibitemShut {NoStop}%
  \bibitem [{\citenamefont {Das}\ \emph {et~al.}(2006)\citenamefont {Das},
    \citenamefont {Sengupta}, \citenamefont {Sen},\ and\ \citenamefont
    {Chakrabarti}}]{das2006infinite}%
    \BibitemOpen
    \bibfield  {author} {\bibinfo {author} {\bibfnamefont {A.}~\bibnamefont
    {Das}}, \bibinfo {author} {\bibfnamefont {K.}~\bibnamefont {Sengupta}},
    \bibinfo {author} {\bibfnamefont {D.}~\bibnamefont {Sen}}, \ and\ \bibinfo
    {author} {\bibfnamefont {B.~K.}\ \bibnamefont {Chakrabarti}},\ }\href
    {\doibase 10.1103/physrevb.74.144423} {\bibfield  {journal} {\bibinfo
    {journal} {Phys. Rev. B}\ }\textbf {\bibinfo {volume} {74}},\ \bibinfo
    {pages} {144423} (\bibinfo {year} {2006})}\BibitemShut {NoStop}%
  \bibitem [{\citenamefont {Britton}\ \emph {et~al.}(2012)\citenamefont
    {Britton}, \citenamefont {Sawyer}, \citenamefont {Keith}, \citenamefont
    {Wang}, \citenamefont {Freericks}, \citenamefont {Uys}, \citenamefont
    {Biercuk},\ and\ \citenamefont {Bollinger}}]{britton2012engineered}%
    \BibitemOpen
    \bibfield  {author} {\bibinfo {author} {\bibfnamefont {J.~W.}\ \bibnamefont
    {Britton}}, \bibinfo {author} {\bibfnamefont {B.~C.}\ \bibnamefont {Sawyer}},
    \bibinfo {author} {\bibfnamefont {A.~C.}\ \bibnamefont {Keith}}, \bibinfo
    {author} {\bibfnamefont {C.-C.~J.}\ \bibnamefont {Wang}}, \bibinfo {author}
    {\bibfnamefont {J.~K.}\ \bibnamefont {Freericks}}, \bibinfo {author}
    {\bibfnamefont {H.}~\bibnamefont {Uys}}, \bibinfo {author} {\bibfnamefont
    {M.~J.}\ \bibnamefont {Biercuk}}, \ and\ \bibinfo {author} {\bibfnamefont
    {J.~J.}\ \bibnamefont {Bollinger}},\ }\href {\doibase 10.1038/nature10981}
    {\bibfield  {journal} {\bibinfo  {journal} {Nature}\ }\textbf {\bibinfo
    {volume} {484}},\ \bibinfo {pages} {489} (\bibinfo {year}
    {2012})}\BibitemShut {NoStop}%
  \bibitem [{\citenamefont {Richerme}\ \emph {et~al.}(2014)\citenamefont
    {Richerme}, \citenamefont {Gong}, \citenamefont {Lee}, \citenamefont {Senko},
    \citenamefont {Smith}, \citenamefont {Foss-Feig}, \citenamefont {Michalakis},
    \citenamefont {Gorshkov},\ and\ \citenamefont {Monroe}}]{richerme2014non}%
    \BibitemOpen
    \bibfield  {author} {\bibinfo {author} {\bibfnamefont {P.}~\bibnamefont
    {Richerme}}, \bibinfo {author} {\bibfnamefont {Z.-X.}\ \bibnamefont {Gong}},
    \bibinfo {author} {\bibfnamefont {A.}~\bibnamefont {Lee}}, \bibinfo {author}
    {\bibfnamefont {C.}~\bibnamefont {Senko}}, \bibinfo {author} {\bibfnamefont
    {J.}~\bibnamefont {Smith}}, \bibinfo {author} {\bibfnamefont
    {M.}~\bibnamefont {Foss-Feig}}, \bibinfo {author} {\bibfnamefont
    {S.}~\bibnamefont {Michalakis}}, \bibinfo {author} {\bibfnamefont {A.~V.}\
    \bibnamefont {Gorshkov}}, \ and\ \bibinfo {author} {\bibfnamefont
    {C.}~\bibnamefont {Monroe}},\ }\href {\doibase 10.1038/nature13450}
    {\bibfield  {journal} {\bibinfo  {journal} {Nature}\ }\textbf {\bibinfo
    {volume} {511}},\ \bibinfo {pages} {198} (\bibinfo {year}
    {2014})}\BibitemShut {NoStop}%
  \bibitem [{\citenamefont {Bohnet}\ \emph {et~al.}(2016)\citenamefont {Bohnet},
    \citenamefont {Sawyer}, \citenamefont {Britton}, \citenamefont {Wall},
    \citenamefont {Rey}, \citenamefont {Foss-Feig},\ and\ \citenamefont
    {Bollinger}}]{bohnet2016quantum}%
    \BibitemOpen
    \bibfield  {author} {\bibinfo {author} {\bibfnamefont {J.~G.}\ \bibnamefont
    {Bohnet}}, \bibinfo {author} {\bibfnamefont {B.~C.}\ \bibnamefont {Sawyer}},
    \bibinfo {author} {\bibfnamefont {J.~W.}\ \bibnamefont {Britton}}, \bibinfo
    {author} {\bibfnamefont {M.~L.}\ \bibnamefont {Wall}}, \bibinfo {author}
    {\bibfnamefont {A.~M.}\ \bibnamefont {Rey}}, \bibinfo {author} {\bibfnamefont
    {M.}~\bibnamefont {Foss-Feig}}, \ and\ \bibinfo {author} {\bibfnamefont
    {J.~J.}\ \bibnamefont {Bollinger}},\ }\href {\doibase
    10.1126/science.aad9958} {\bibfield  {journal} {\bibinfo  {journal}
    {Science}\ }\textbf {\bibinfo {volume} {352}},\ \bibinfo {pages} {1297}
    (\bibinfo {year} {2016})}\BibitemShut {NoStop}%
  \bibitem [{\citenamefont {Zibold}\ \emph {et~al.}(2010)\citenamefont {Zibold},
    \citenamefont {Nicklas}, \citenamefont {Gross},\ and\ \citenamefont
    {Oberthaler}}]{zibold2010classical}%
    \BibitemOpen
    \bibfield  {author} {\bibinfo {author} {\bibfnamefont {T.}~\bibnamefont
    {Zibold}}, \bibinfo {author} {\bibfnamefont {E.}~\bibnamefont {Nicklas}},
    \bibinfo {author} {\bibfnamefont {C.}~\bibnamefont {Gross}}, \ and\ \bibinfo
    {author} {\bibfnamefont {M.~K.}\ \bibnamefont {Oberthaler}},\ }\href
    {\doibase 10.1103/physrevlett.105.204101} {\bibfield  {journal} {\bibinfo
    {journal} {Phys. Rev. Lett.}\ }\textbf {\bibinfo {volume} {105}},\ \bibinfo
    {pages} {204101} (\bibinfo {year} {2010})}\BibitemShut {NoStop}%
  \bibitem [{\citenamefont {Est{\`{e}}ve}\ \emph {et~al.}(2008)\citenamefont
    {Est{\`{e}}ve}, \citenamefont {Gross}, \citenamefont {Weller}, \citenamefont
    {Giovanazzi},\ and\ \citenamefont {Oberthaler}}]{esteve2008squeezing}%
    \BibitemOpen
    \bibfield  {author} {\bibinfo {author} {\bibfnamefont {J.}~\bibnamefont
    {Est{\`{e}}ve}}, \bibinfo {author} {\bibfnamefont {C.}~\bibnamefont {Gross}},
    \bibinfo {author} {\bibfnamefont {A.}~\bibnamefont {Weller}}, \bibinfo
    {author} {\bibfnamefont {S.}~\bibnamefont {Giovanazzi}}, \ and\ \bibinfo
    {author} {\bibfnamefont {M.~K.}\ \bibnamefont {Oberthaler}},\ }\href
    {\doibase 10.1038/nature07332} {\bibfield  {journal} {\bibinfo  {journal}
    {Nature}\ }\textbf {\bibinfo {volume} {455}},\ \bibinfo {pages} {1216}
    (\bibinfo {year} {2008})}\BibitemShut {NoStop}%
  \bibitem [{\citenamefont {Gross}\ \emph {et~al.}(2010)\citenamefont {Gross},
    \citenamefont {Zibold}, \citenamefont {Nicklas}, \citenamefont
    {Est{\`{e}}ve},\ and\ \citenamefont {Oberthaler}}]{gross2010nonlinear}%
    \BibitemOpen
    \bibfield  {author} {\bibinfo {author} {\bibfnamefont {C.}~\bibnamefont
    {Gross}}, \bibinfo {author} {\bibfnamefont {T.}~\bibnamefont {Zibold}},
    \bibinfo {author} {\bibfnamefont {E.}~\bibnamefont {Nicklas}}, \bibinfo
    {author} {\bibfnamefont {J.}~\bibnamefont {Est{\`{e}}ve}}, \ and\ \bibinfo
    {author} {\bibfnamefont {M.~K.}\ \bibnamefont {Oberthaler}},\ }\href
    {\doibase 10.1038/nature08919} {\bibfield  {journal} {\bibinfo  {journal}
    {Nature}\ }\textbf {\bibinfo {volume} {464}},\ \bibinfo {pages} {1165}
    (\bibinfo {year} {2010})}\BibitemShut {NoStop}%
  \bibitem [{\citenamefont {Juli{\'{a}}-D{\'{\i}}az}\ \emph
    {et~al.}(2012)\citenamefont {Juli{\'{a}}-D{\'{\i}}az}, \citenamefont
    {Zibold}, \citenamefont {Oberthaler}, \citenamefont {Mel{\'{e}}-Messeguer},
    \citenamefont {Martorell},\ and\ \citenamefont
    {Polls}}]{julia-diaz2012dynamic}%
    \BibitemOpen
    \bibfield  {author} {\bibinfo {author} {\bibfnamefont {B.}~\bibnamefont
    {Juli{\'{a}}-D{\'{\i}}az}}, \bibinfo {author} {\bibfnamefont
    {T.}~\bibnamefont {Zibold}}, \bibinfo {author} {\bibfnamefont {M.~K.}\
    \bibnamefont {Oberthaler}}, \bibinfo {author} {\bibfnamefont
    {M.}~\bibnamefont {Mel{\'{e}}-Messeguer}}, \bibinfo {author} {\bibfnamefont
    {J.}~\bibnamefont {Martorell}}, \ and\ \bibinfo {author} {\bibfnamefont
    {A.}~\bibnamefont {Polls}},\ }\href {\doibase 10.1103/physreva.86.023615}
    {\bibfield  {journal} {\bibinfo  {journal} {Phys. Rev. A}\ }\textbf {\bibinfo
    {volume} {86}},\ \bibinfo {pages} {023615} (\bibinfo {year}
    {2012})}\BibitemShut {NoStop}%
  \bibitem [{\citenamefont {Haroutyunyan}\ and\ \citenamefont
    {Nienhuis}(2003)}]{haroutyunyan2003analogy}%
    \BibitemOpen
    \bibfield  {author} {\bibinfo {author} {\bibfnamefont {H.~L.}\ \bibnamefont
    {Haroutyunyan}}\ and\ \bibinfo {author} {\bibfnamefont {G.}~\bibnamefont
    {Nienhuis}},\ }\href {\doibase 10.1103/physreva.67.053611} {\bibfield
    {journal} {\bibinfo  {journal} {Physical Review A}\ }\textbf {\bibinfo
    {volume} {67}},\ \bibinfo {pages} {053611} (\bibinfo {year}
    {2003})}\BibitemShut {NoStop}%
  \bibitem [{\citenamefont {Krahn}\ and\ \citenamefont
    {O'Dell}(2009)}]{krahn2009classical}%
    \BibitemOpen
    \bibfield  {author} {\bibinfo {author} {\bibfnamefont {G.~J.}\ \bibnamefont
    {Krahn}}\ and\ \bibinfo {author} {\bibfnamefont {D.~H.~J.}\ \bibnamefont
    {O'Dell}},\ }\href {\doibase 10.1088/0953-4075/42/20/205501} {\bibfield
    {journal} {\bibinfo  {journal} {Journal of Physics B: Atomic, Molecular and
    Optical Physics}\ }\textbf {\bibinfo {volume} {42}},\ \bibinfo {pages}
    {205501} (\bibinfo {year} {2009})}\BibitemShut {NoStop}%
  \bibitem [{\citenamefont {Veksler}\ and\ \citenamefont
    {Fishman}(2015)}]{veksler2015semiclassical}%
    \BibitemOpen
    \bibfield  {author} {\bibinfo {author} {\bibfnamefont {H.}~\bibnamefont
    {Veksler}}\ and\ \bibinfo {author} {\bibfnamefont {S.}~\bibnamefont
    {Fishman}},\ }\href {\doibase 10.1088/1367-2630/17/5/053030} {\bibfield
    {journal} {\bibinfo  {journal} {New J. Phys.}\ }\textbf {\bibinfo {volume}
    {17}},\ \bibinfo {pages} {053030} (\bibinfo {year} {2015})}\BibitemShut
    {NoStop}%
  \bibitem [{\citenamefont {Mumford}\ \emph {et~al.}(2019)\citenamefont
    {Mumford}, \citenamefont {Turner}, \citenamefont {Sprung},\ and\
    \citenamefont {O'Dell}}]{mumford2019quantum}%
    \BibitemOpen
    \bibfield  {author} {\bibinfo {author} {\bibfnamefont {J.}~\bibnamefont
    {Mumford}}, \bibinfo {author} {\bibfnamefont {E.}~\bibnamefont {Turner}},
    \bibinfo {author} {\bibfnamefont {D.~W.~L.}\ \bibnamefont {Sprung}}, \ and\
    \bibinfo {author} {\bibfnamefont {D.~H.~J.}\ \bibnamefont {O'Dell}},\ }\href
    {\doibase 10.1103/physrevlett.122.170402} {\bibfield  {journal} {\bibinfo
    {journal} {Phys. Rev. Lett.}\ }\textbf {\bibinfo {volume} {122}},\ \bibinfo
    {pages} {170402} (\bibinfo {year} {2019})}\BibitemShut {NoStop}%
  \bibitem [{\citenamefont {Chin}\ \emph {et~al.}(2010)\citenamefont {Chin},
    \citenamefont {Grimm}, \citenamefont {Julienne},\ and\ \citenamefont
    {Tiesinga}}]{chin2010feshbach}%
    \BibitemOpen
    \bibfield  {author} {\bibinfo {author} {\bibfnamefont {C.}~\bibnamefont
    {Chin}}, \bibinfo {author} {\bibfnamefont {R.}~\bibnamefont {Grimm}},
    \bibinfo {author} {\bibfnamefont {P.}~\bibnamefont {Julienne}}, \ and\
    \bibinfo {author} {\bibfnamefont {E.}~\bibnamefont {Tiesinga}},\ }\href
    {\doibase 10.1103/RevModPhys.82.1225} {\bibfield  {journal} {\bibinfo
    {journal} {Rev. Mod. Phys.}\ }\textbf {\bibinfo {volume} {82}},\ \bibinfo
    {pages} {1225} (\bibinfo {year} {2010})}\BibitemShut {NoStop}%
  \bibitem [{\citenamefont {Susskind}\ and\ \citenamefont
    {Glogower}(1964)}]{susskind1964quantum}%
    \BibitemOpen
    \bibfield  {author} {\bibinfo {author} {\bibfnamefont {L.}~\bibnamefont
    {Susskind}}\ and\ \bibinfo {author} {\bibfnamefont {J.}~\bibnamefont
    {Glogower}},\ }\href {\doibase 10.1103/PhysicsPhysiqueFizika.1.49} {\bibfield
     {journal} {\bibinfo  {journal} {Physics}\ }\textbf {\bibinfo {volume} {1}},\
    \bibinfo {pages} {49} (\bibinfo {year} {1964})}\BibitemShut {NoStop}%
  \bibitem [{\citenamefont {Carruthers}\ and\ \citenamefont
    {Nieto}(1968)}]{carruthers1968phase}%
    \BibitemOpen
    \bibfield  {author} {\bibinfo {author} {\bibfnamefont {P.}~\bibnamefont
    {Carruthers}}\ and\ \bibinfo {author} {\bibfnamefont {M.~M.}\ \bibnamefont
    {Nieto}},\ }\href {\doibase 10.1103/RevModPhys.40.411} {\bibfield  {journal}
    {\bibinfo  {journal} {Rev. Mod. Phys.}\ }\textbf {\bibinfo {volume} {40}},\
    \bibinfo {pages} {411} (\bibinfo {year} {1968})}\BibitemShut {NoStop}%
  \bibitem [{\citenamefont {Pegg}\ and\ \citenamefont
    {Barnett}(1989)}]{pegg1989phase}%
    \BibitemOpen
    \bibfield  {author} {\bibinfo {author} {\bibfnamefont {D.~T.}\ \bibnamefont
    {Pegg}}\ and\ \bibinfo {author} {\bibfnamefont {S.~M.}\ \bibnamefont
    {Barnett}},\ }\href {\doibase 10.1103/PhysRevA.39.1665} {\bibfield  {journal}
    {\bibinfo  {journal} {Phys. Rev. A}\ }\textbf {\bibinfo {volume} {39}},\
    \bibinfo {pages} {1665} (\bibinfo {year} {1989})}\BibitemShut {NoStop}%
  \bibitem [{\citenamefont {Alonso}\ and\ \citenamefont
    {Dennis}(2017)}]{alonso2017ray}%
    \BibitemOpen
    \bibfield  {author} {\bibinfo {author} {\bibfnamefont {M.~A.}\ \bibnamefont
    {Alonso}}\ and\ \bibinfo {author} {\bibfnamefont {M.~R.}\ \bibnamefont
    {Dennis}},\ }\href {\doibase 10.1364/optica.4.000476} {\bibfield  {journal}
    {\bibinfo  {journal} {Optica}\ }\textbf {\bibinfo {volume} {4}},\ \bibinfo
    {pages} {476} (\bibinfo {year} {2017})}\BibitemShut {NoStop}%
  \bibitem [{\citenamefont {Shchesnovich}\ and\ \citenamefont
    {Trippenbach}(2008)}]{shchesnovich2008fock}%
    \BibitemOpen
    \bibfield  {author} {\bibinfo {author} {\bibfnamefont {V.~S.}\ \bibnamefont
    {Shchesnovich}}\ and\ \bibinfo {author} {\bibfnamefont {M.}~\bibnamefont
    {Trippenbach}},\ }\href {\doibase 10.1103/PhysRevA.78.023611} {\bibfield
    {journal} {\bibinfo  {journal} {Phys. Rev. A}\ }\textbf {\bibinfo {volume}
    {78}},\ \bibinfo {pages} {023611} (\bibinfo {year} {2008})}\BibitemShut
    {NoStop}%
  \bibitem [{\citenamefont {Zor}\ and\ \citenamefont
    {Kay}(1996)}]{zor1996globally}%
    \BibitemOpen
    \bibfield  {author} {\bibinfo {author} {\bibfnamefont {D.}~\bibnamefont
    {Zor}}\ and\ \bibinfo {author} {\bibfnamefont {K.~G.}\ \bibnamefont {Kay}},\
    }\href {\doibase 10.1103/physrevlett.76.1990} {\bibfield  {journal} {\bibinfo
     {journal} {Phys. Rev. Lett.}\ }\textbf {\bibinfo {volume} {76}},\ \bibinfo
    {pages} {1990} (\bibinfo {year} {1996})}\BibitemShut {NoStop}%
  \bibitem [{\citenamefont {Forbes}\ and\ \citenamefont
    {Alonso}(2001)}]{forbes2001using}%
    \BibitemOpen
    \bibfield  {author} {\bibinfo {author} {\bibfnamefont {G.~W.}\ \bibnamefont
    {Forbes}}\ and\ \bibinfo {author} {\bibfnamefont {M.~A.}\ \bibnamefont
    {Alonso}},\ }\href {\doibase 10.1364/josaa.18.001132} {\bibfield  {journal}
    {\bibinfo  {journal} {J. Opt. Soc. Am. A}\ }\textbf {\bibinfo {volume}
    {18}},\ \bibinfo {pages} {1132} (\bibinfo {year} {2001})}\BibitemShut
    {NoStop}%
  \bibitem [{\citenamefont {Alonso}\ and\ \citenamefont
    {Forbes}(2001)}]{alonso2001using}%
    \BibitemOpen
    \bibfield  {author} {\bibinfo {author} {\bibfnamefont {M.~A.}\ \bibnamefont
    {Alonso}}\ and\ \bibinfo {author} {\bibfnamefont {G.~W.}\ \bibnamefont
    {Forbes}},\ }\href {\doibase 10.1364/josaa.18.001146} {\bibfield  {journal}
    {\bibinfo  {journal} {J. Opt. Soc. Am. A}\ }\textbf {\bibinfo {volume}
    {18}},\ \bibinfo {pages} {1146} (\bibinfo {year} {2001})}\BibitemShut
    {NoStop}%
  \bibitem [{\citenamefont {Alonso}\ and\ \citenamefont
    {Forbes}(2002)}]{alonso2002stable}%
    \BibitemOpen
    \bibfield  {author} {\bibinfo {author} {\bibfnamefont {M.~A.}\ \bibnamefont
    {Alonso}}\ and\ \bibinfo {author} {\bibfnamefont {G.~W.}\ \bibnamefont
    {Forbes}},\ }\href {\doibase 10.1364/oe.10.000728} {\bibfield  {journal}
    {\bibinfo  {journal} {Opt. Express}\ }\textbf {\bibinfo {volume} {10}},\
    \bibinfo {pages} {728} (\bibinfo {year} {2002})}\BibitemShut {NoStop}%
  \bibitem [{\citenamefont {Pancharatnam}(1956)}]{pancharatnam1956generalized}%
    \BibitemOpen
    \bibfield  {author} {\bibinfo {author} {\bibfnamefont {S.}~\bibnamefont
    {Pancharatnam}},\ }in\ \href@noop {} {\emph {\bibinfo {booktitle} {Proc. Ind.
    Acad. Sci.}}},\ Vol.~\bibinfo {volume} {44}\ (\bibinfo {organization}
    {Springer},\ \bibinfo {year} {1956})\ pp.\ \bibinfo {pages}
    {247--262}\BibitemShut {NoStop}%
  \bibitem [{\citenamefont {Berry}(1987)}]{berry1987adiabatic}%
    \BibitemOpen
    \bibfield  {author} {\bibinfo {author} {\bibfnamefont {M.~V.}\ \bibnamefont
    {Berry}},\ }\href {\doibase 10.1080/09500348714551321} {\bibfield  {journal}
    {\bibinfo  {journal} {J. Mod. Opt.}\ }\textbf {\bibinfo {volume} {34}},\
    \bibinfo {pages} {1401} (\bibinfo {year} {1987})}\BibitemShut {NoStop}%
  \bibitem [{\citenamefont {Malhotra}\ \emph {et~al.}(2018)\citenamefont
    {Malhotra}, \citenamefont {Guti{\'{e}}rrez-Cuevas}, \citenamefont {Hassett},
    \citenamefont {Dennis}, \citenamefont {Vamivakas},\ and\ \citenamefont
    {Alonso}}]{malhotra2018measuring}%
    \BibitemOpen
    \bibfield  {author} {\bibinfo {author} {\bibfnamefont {T.}~\bibnamefont
    {Malhotra}}, \bibinfo {author} {\bibfnamefont {R.}~\bibnamefont
    {Guti{\'{e}}rrez-Cuevas}}, \bibinfo {author} {\bibfnamefont {J.}~\bibnamefont
    {Hassett}}, \bibinfo {author} {\bibfnamefont {M.~R.}\ \bibnamefont {Dennis}},
    \bibinfo {author} {\bibfnamefont {A.~N.}\ \bibnamefont {Vamivakas}}, \ and\
    \bibinfo {author} {\bibfnamefont {M.~A.}\ \bibnamefont {Alonso}},\ }\href
    {\doibase 10.1103/physrevlett.120.233602} {\bibfield  {journal} {\bibinfo
    {journal} {Phys. Rev. Lett.}\ }\textbf {\bibinfo {volume} {120}},\ \bibinfo
    {pages} {233602} (\bibinfo {year} {2018})}\BibitemShut {NoStop}%
  \bibitem [{\citenamefont {Shen}\ \emph {et~al.}(2020)\citenamefont {Shen},
    \citenamefont {Yang}, \citenamefont {Naidoo}, \citenamefont {Fu},\ and\
    \citenamefont {Forbes}}]{shen2020structured}%
    \BibitemOpen
    \bibfield  {author} {\bibinfo {author} {\bibfnamefont {Y.}~\bibnamefont
    {Shen}}, \bibinfo {author} {\bibfnamefont {X.}~\bibnamefont {Yang}}, \bibinfo
    {author} {\bibfnamefont {D.}~\bibnamefont {Naidoo}}, \bibinfo {author}
    {\bibfnamefont {X.}~\bibnamefont {Fu}}, \ and\ \bibinfo {author}
    {\bibfnamefont {A.}~\bibnamefont {Forbes}},\ }\href {\doibase
    10.1364/optica.382994} {\bibfield  {journal} {\bibinfo  {journal} {Optica}\
    }\textbf {\bibinfo {volume} {7}},\ \bibinfo {pages} {820} (\bibinfo {year}
    {2020})}\BibitemShut {NoStop}%
  \bibitem [{\citenamefont {Jaffe}\ \emph {et~al.}(2021)\citenamefont {Jaffe},
    \citenamefont {Palm}, \citenamefont {Baum}, \citenamefont {Taneja},\ and\
    \citenamefont {Simon}}]{jaffe2021aberrated}%
    \BibitemOpen
    \bibfield  {author} {\bibinfo {author} {\bibfnamefont {M.}~\bibnamefont
    {Jaffe}}, \bibinfo {author} {\bibfnamefont {L.}~\bibnamefont {Palm}},
    \bibinfo {author} {\bibfnamefont {C.}~\bibnamefont {Baum}}, \bibinfo {author}
    {\bibfnamefont {L.}~\bibnamefont {Taneja}}, \ and\ \bibinfo {author}
    {\bibfnamefont {J.}~\bibnamefont {Simon}},\ }\href {\doibase
    10.1103/physreva.104.013524} {\ \textbf {\bibinfo {volume} {104}} (\bibinfo
    {year} {2021}),\ 10.1103/physreva.104.013524}\BibitemShut {NoStop}%
  \bibitem [{\citenamefont {Schine}\ \emph {et~al.}(2016)\citenamefont {Schine},
    \citenamefont {Ryou}, \citenamefont {Gromov}, \citenamefont {Sommer},\ and\
    \citenamefont {Simon}}]{schine2016synthetic}%
    \BibitemOpen
    \bibfield  {author} {\bibinfo {author} {\bibfnamefont {N.}~\bibnamefont
    {Schine}}, \bibinfo {author} {\bibfnamefont {A.}~\bibnamefont {Ryou}},
    \bibinfo {author} {\bibfnamefont {A.}~\bibnamefont {Gromov}}, \bibinfo
    {author} {\bibfnamefont {A.}~\bibnamefont {Sommer}}, \ and\ \bibinfo {author}
    {\bibfnamefont {J.}~\bibnamefont {Simon}},\ }\href {\doibase
    10.1038/nature17943} {\bibfield  {journal} {\bibinfo  {journal} {Nature}\
    }\textbf {\bibinfo {volume} {534}},\ \bibinfo {pages} {671} (\bibinfo {year}
    {2016})}\BibitemShut {NoStop}%
  \bibitem [{\citenamefont {Clark}\ \emph {et~al.}(2020)\citenamefont {Clark},
    \citenamefont {Schine}, \citenamefont {Baum}, \citenamefont {Jia},\ and\
    \citenamefont {Simon}}]{clark2020observation}%
    \BibitemOpen
    \bibfield  {author} {\bibinfo {author} {\bibfnamefont {L.~W.}\ \bibnamefont
    {Clark}}, \bibinfo {author} {\bibfnamefont {N.}~\bibnamefont {Schine}},
    \bibinfo {author} {\bibfnamefont {C.}~\bibnamefont {Baum}}, \bibinfo {author}
    {\bibfnamefont {N.}~\bibnamefont {Jia}}, \ and\ \bibinfo {author}
    {\bibfnamefont {J.}~\bibnamefont {Simon}},\ }\href {\doibase
    10.1038/s41586-020-2318-5} {\bibfield  {journal} {\bibinfo  {journal}
    {Nature}\ }\textbf {\bibinfo {volume} {582}},\ \bibinfo {pages} {41}
    (\bibinfo {year} {2020})}\BibitemShut {NoStop}%
  \bibitem [{\citenamefont {Dennis}(2006)}]{dennis2006rows}%
    \BibitemOpen
    \bibfield  {author} {\bibinfo {author} {\bibfnamefont {M.~R.}\ \bibnamefont
    {Dennis}},\ }\href {\doibase 10.1364/ol.31.001325} {\bibfield  {journal}
    {\bibinfo  {journal} {Opt. Lett.}\ }\textbf {\bibinfo {volume} {31}},\
    \bibinfo {pages} {1325} (\bibinfo {year} {2006})}\BibitemShut {NoStop}%
  \bibitem [{\citenamefont {lin Chao}(1974)}]{chao1974high}%
    \BibitemOpen
    \bibfield  {author} {\bibinfo {author} {\bibfnamefont {S.}~\bibnamefont {lin
    Chao}},\ }\emph {\bibinfo {title} {High order transverse modes in an
    astigmatic cavity}},\ \href@noop {} {Ph.D. thesis},\ \bibinfo  {school}
    {University of Rochester} (\bibinfo {year} {1974})\BibitemShut {NoStop}%
  \bibitem [{\citenamefont {Ulyanov}\ and\ \citenamefont
    {Zaslavskii}(1992)}]{ulyanov1992new}%
    \BibitemOpen
    \bibfield  {author} {\bibinfo {author} {\bibfnamefont {V.~V.}\ \bibnamefont
    {Ulyanov}}\ and\ \bibinfo {author} {\bibfnamefont {O.~B.}\ \bibnamefont
    {Zaslavskii}},\ }\href {\doibase 10.1016/0370-1573(92)90158-V} {\bibfield
    {journal} {\bibinfo  {journal} {Phys. Rep.}\ }\textbf {\bibinfo {volume}
    {216}},\ \bibinfo {pages} {179} (\bibinfo {year} {1992})}\BibitemShut
    {NoStop}%
  \bibitem [{\citenamefont {Mumford}\ \emph {et~al.}(2017)\citenamefont
    {Mumford}, \citenamefont {Kirkby},\ and\ \citenamefont
    {O'Dell}}]{mumford2017catastrophes}%
    \BibitemOpen
    \bibfield  {author} {\bibinfo {author} {\bibfnamefont {J.}~\bibnamefont
    {Mumford}}, \bibinfo {author} {\bibfnamefont {W.}~\bibnamefont {Kirkby}}, \
    and\ \bibinfo {author} {\bibfnamefont {D.~H.~J.}\ \bibnamefont {O'Dell}},\
    }\href {\doibase 10.1088/1361-6455/aa56af} {\bibfield  {journal} {\bibinfo
    {journal} {J. Phys. B: At., Mol. Opt. Phys.}\ }\textbf {\bibinfo {volume}
    {50}},\ \bibinfo {pages} {044005} (\bibinfo {year} {2017})}\BibitemShut
    {NoStop}%
  \bibitem [{\citenamefont {Vourdas}(1990)}]{vourdas1990su2}%
    \BibitemOpen
    \bibfield  {author} {\bibinfo {author} {\bibfnamefont {A.}~\bibnamefont
    {Vourdas}},\ }\href {\doibase 10.1103/PhysRevA.41.1653} {\bibfield  {journal}
    {\bibinfo  {journal} {Phys. Rev. A}\ }\textbf {\bibinfo {volume} {41}},\
    \bibinfo {pages} {1653} (\bibinfo {year} {1990})}\BibitemShut {NoStop}%
  \bibitem [{\citenamefont {Luis}\ and\ \citenamefont
    {S\'{a}nchez-Soto}(1993)}]{luis1993phase}%
    \BibitemOpen
    \bibfield  {author} {\bibinfo {author} {\bibfnamefont {A.}~\bibnamefont
    {Luis}}\ and\ \bibinfo {author} {\bibfnamefont {L.~L.}\ \bibnamefont
    {S\'{a}nchez-Soto}},\ }\href {\doibase 10.1103/PhysRevA.48.4702} {\bibfield
    {journal} {\bibinfo  {journal} {Phys. Rev. A}\ }\textbf {\bibinfo {volume}
    {48}},\ \bibinfo {pages} {4702} (\bibinfo {year} {1993})}\BibitemShut
    {NoStop}%
  \bibitem [{\citenamefont {Longhi}(2011)}]{longhi2011optical}%
    \BibitemOpen
    \bibfield  {author} {\bibinfo {author} {\bibfnamefont {S.}~\bibnamefont
    {Longhi}},\ }\href {\doibase 10.1088/0953-4075/44/5/051001} {\bibfield
    {journal} {\bibinfo  {journal} {J. Phys. B: At., Mol. Opt. Phys.}\ }\textbf
    {\bibinfo {volume} {44}},\ \bibinfo {pages} {051001} (\bibinfo {year}
    {2011})}\BibitemShut {NoStop}%
  \bibitem [{\citenamefont {Morales}\ \emph {et~al.}(2017)\citenamefont
    {Morales}, \citenamefont {Rodr{\'{\i}}guez-Lara},\ and\ \citenamefont
    {Malomed}}]{morales2017polarization}%
    \BibitemOpen
    \bibfield  {author} {\bibinfo {author} {\bibfnamefont {J.~D.~H.}\
    \bibnamefont {Morales}}, \bibinfo {author} {\bibfnamefont {B.~M.}\
    \bibnamefont {Rodr{\'{\i}}guez-Lara}}, \ and\ \bibinfo {author}
    {\bibfnamefont {B.~A.}\ \bibnamefont {Malomed}},\ }\href {\doibase
    10.1364/ol.42.004402} {\bibfield  {journal} {\bibinfo  {journal} {Opt.
    Lett.}\ }\textbf {\bibinfo {volume} {42}},\ \bibinfo {pages} {4402} (\bibinfo
    {year} {2017})}\BibitemShut {NoStop}%
  \bibitem [{\citenamefont {Becker}\ \emph {et~al.}(2017)\citenamefont {Becker},
    \citenamefont {Mirahmadi}, \citenamefont {Schmidt}, \citenamefont {Schatz},\
    and\ \citenamefont {Friedrich}}]{becker2017conditional}%
    \BibitemOpen
    \bibfield  {author} {\bibinfo {author} {\bibfnamefont {S.}~\bibnamefont
    {Becker}}, \bibinfo {author} {\bibfnamefont {M.}~\bibnamefont {Mirahmadi}},
    \bibinfo {author} {\bibfnamefont {B.}~\bibnamefont {Schmidt}}, \bibinfo
    {author} {\bibfnamefont {K.}~\bibnamefont {Schatz}}, \ and\ \bibinfo {author}
    {\bibfnamefont {B.}~\bibnamefont {Friedrich}},\ }\href {\doibase
    10.1140/epjd/e2017-80134-6} {\bibfield  {journal} {\bibinfo  {journal} {Eur.
    Phys. J. D}\ }\textbf {\bibinfo {volume} {71}},\ \bibinfo {pages} {149}
    (\bibinfo {year} {2017})}\BibitemShut {NoStop}%
  \bibitem [{\citenamefont {Razavy}(1980)}]{razavy1980exactly}%
    \BibitemOpen
    \bibfield  {author} {\bibinfo {author} {\bibfnamefont {M.}~\bibnamefont
    {Razavy}},\ }\href {\doibase 10.1119/1.12141} {\bibfield  {journal} {\bibinfo
     {journal} {Am. J. Phys}\ }\textbf {\bibinfo {volume} {48}},\ \bibinfo
    {pages} {285} (\bibinfo {year} {1980})}\BibitemShut {NoStop}%
  \bibitem [{\citenamefont {Colombo}(1966)}]{colombo1966cassinis}%
    \BibitemOpen
    \bibfield  {author} {\bibinfo {author} {\bibfnamefont {G.}~\bibnamefont
    {Colombo}},\ }\href {\doibase 10.1086/109983} {\bibfield  {journal} {\bibinfo
     {journal} {Astron. J.}\ }\textbf {\bibinfo {volume} {71}},\ \bibinfo {pages}
    {891} (\bibinfo {year} {1966})}\BibitemShut {NoStop}%
  \bibitem [{\citenamefont {Henrard}\ and\ \citenamefont
    {Murigande}(1987)}]{henrard1987colombos}%
    \BibitemOpen
    \bibfield  {author} {\bibinfo {author} {\bibfnamefont {J.}~\bibnamefont
    {Henrard}}\ and\ \bibinfo {author} {\bibfnamefont {C.}~\bibnamefont
    {Murigande}},\ }\href {\doibase 10.1007/bf01235852} {\bibfield  {journal}
    {\bibinfo  {journal} {Celestial Mech}\ }\textbf {\bibinfo {volume} {40}},\
    \bibinfo {pages} {345} (\bibinfo {year} {1987})}\BibitemShut {NoStop}%
  \bibitem [{\citenamefont {Law}\ \emph {et~al.}(1998)\citenamefont {Law},
    \citenamefont {Pu},\ and\ \citenamefont {Bigelow}}]{law1998quantum}%
    \BibitemOpen
    \bibfield  {author} {\bibinfo {author} {\bibfnamefont {C.~K.}\ \bibnamefont
    {Law}}, \bibinfo {author} {\bibfnamefont {H.}~\bibnamefont {Pu}}, \ and\
    \bibinfo {author} {\bibfnamefont {N.~P.}\ \bibnamefont {Bigelow}},\ }\href
    {\doibase 10.1103/physrevlett.81.5257} {\bibfield  {journal} {\bibinfo
    {journal} {Phys Rev Lett}\ }\textbf {\bibinfo {volume} {81}},\ \bibinfo
    {pages} {5257} (\bibinfo {year} {1998})}\BibitemShut {NoStop}%
  \bibitem [{\citenamefont {Evrard}\ \emph {et~al.}(2021)\citenamefont {Evrard},
    \citenamefont {Qu}, \citenamefont {Dalibard},\ and\ \citenamefont
    {Gerbier}}]{evrard2021many}%
    \BibitemOpen
    \bibfield  {author} {\bibinfo {author} {\bibfnamefont {B.}~\bibnamefont
    {Evrard}}, \bibinfo {author} {\bibfnamefont {A.}~\bibnamefont {Qu}}, \bibinfo
    {author} {\bibfnamefont {J.}~\bibnamefont {Dalibard}}, \ and\ \bibinfo
    {author} {\bibfnamefont {F.}~\bibnamefont {Gerbier}},\ }\href {\doibase
    10.1103/physrevlett.126.063401} {\bibfield  {journal} {\bibinfo  {journal}
    {Phys Rev Lett}\ }\textbf {\bibinfo {volume} {126}},\ \bibinfo {pages}
    {063401} (\bibinfo {year} {2021})}\BibitemShut {NoStop}%
  \bibitem [{\citenamefont {Kirkby}\ \emph {et~al.}(2022)\citenamefont {Kirkby},
    \citenamefont {Yee}, \citenamefont {Shi},\ and\ \citenamefont
    {O{\textquotesingle}Dell}}]{kirkby2022caustics}%
    \BibitemOpen
    \bibfield  {author} {\bibinfo {author} {\bibfnamefont {W.}~\bibnamefont
    {Kirkby}}, \bibinfo {author} {\bibfnamefont {Y.}~\bibnamefont {Yee}},
    \bibinfo {author} {\bibfnamefont {K.}~\bibnamefont {Shi}}, \ and\ \bibinfo
    {author} {\bibfnamefont {D.~H.~J.}\ \bibnamefont {O{\textquotesingle}Dell}},\
    }\href {\doibase 10.1103/physrevresearch.4.013105} {\bibfield  {journal}
    {\bibinfo  {journal} {Phys. Rev. Research}\ }\textbf {\bibinfo {volume}
    {4}},\ \bibinfo {pages} {013105} (\bibinfo {year} {2022})}\BibitemShut
    {NoStop}%
  \end{thebibliography}

%

\end{document}